\documentclass[12pt]{article}

\usepackage[english]{babel}
\usepackage[latin1]{inputenc}
\usepackage{graphicx}
\usepackage{amsfonts}
\usepackage{color}
\usepackage{commath}
\usepackage{verbatim}
\usepackage{amsmath}
\usepackage{mathtools}
\usepackage{array}

\begin{document}

\begin{titlepage}

\begin{flushright}
%arXiv:MMYY.XXXXX
\end{flushright}
\vskip 2.5cm

\begin{center}
{\Large Solution to the $\beta$-functions in Lorentz-violating theories as a decomposition into irreducible representations}
\end{center}

\vspace{1ex}

\begin{center}
{\large Alejandro Ferrero\footnote{{\tt a.ferrero@uniandes.edu.co}}}

\vspace{5mm}
{\sl Departmento de F\'isica} \\
{\sl Universidad de los Andes} \\
{\sl Bogot\'a, Colombia} \\

\end{center}

\vspace{2.5ex}

\medskip

\centerline{\bf Abstract}

\bigskip

We analyze the $\beta$-functions of Yukawa and electromagnetic theories with Lorentz violation (LV) and propose an alternative method to find the scale dependence of the different fields that parameterize such violations. The method of solution consists of decomposing a family of parameters into their irreducible representations and thus generating a group of subfamilies that obey the same symmetries and transformation rules. The convenience of this method relies on the fact that the degree of complexity of the problem is reduced in the decoupling process. For a set of parameters describing a Lorentz-violating theory, we expect their associated $\beta$-functions to be nonnegative or, otherwise, their scale dependence weak enough. These conditions are necessary because asymptotically-free parameters could leave high imprints of LV at low energies, which are ruled out by observations. Besides imposing some constrains on the coefficients that describe LV, this method can be used to find the fixed points of the theory in each particular irreducible subspace.

\bigskip

\end{titlepage}

\newpage

\section{Introduction}\label{sec:intro}

Although LV has not been observed in our relatively low-energy experiments, there is no reason to assume that small violations of Lorentz and CPT symmetries are incompatible with quantum field theories \cite{ref-rev1,ref-rev2}. Nonetheless, any model that includes Lorentz or CPT violations, for instance, the Standard Model Extension (SME) \cite{ref-rev1,ref-rev2,ref-kost2}, must be compatible with the phenomena observed in our universe.

The conservation of Lorentz and CPT symmetries should be related. In 2002, Greenberg proposed that CPT violations imply the violation of Lorentz invariance \cite{ref-greenberg}. Nevertheless, Deutsch and Gracia-Bond\'ia recently suggested that such claim is still on somewhat shaky ground \cite{ref-pepe}. While the relation between both symmetries might still be unclear, there are some fundamental causes that could induce their respective violations. Some possible scenarios include models with time-variating coupling constants \cite{ref-ferrero1,ref-ferrero2,ref-webb1,ref-webb2,ref-jean,ref-gould} and models with noncommutative geometry, spacetime discretization, among others \cite{ref-carroll,ref-gambini,ref-k-sam,ref-singh}.

The values of some of the parameters described by the SME can be constrained by means of experiments. Most of the analysis in the SME has been performed in the electromagnetic sector \cite{ref-stecker,ref-tobar,ref-altschul20,ref-boquet,ref-altschul22}. Additionally, some work has been performed in the gravitational \cite{ref-quentin,ref-zhao} and neutrino regimes \cite{ref-miniboone}. A possible violation of Lorentz symmetry in the Yukawa sector has not been strongly emphasized in spite of its great importance \cite{ref-Anderson,ref-anber}, specially with the great evidence supporting the existence of the Higgs particle coming from the LHC experiments \cite{ref-higgs1}. The existence of the Higgs particle, which generates the masses of the particles in the standard model, opens the window to have a better understanding of Yukawa interactions.

Besides using observations to test the validity of models that include Lorentz and CPT violations, the renormalization group and the scale dependence of the parameters that describe such violations are important to formulate a consistent theory. As it is well known, the scale-dependence of a physical quantity in a quantum field theory is described by its $\beta$-function. While a positive value of such function implies that the observed value of its associated physical quantity increases with the energy scale and a negative value describes the opposite behavior, the zeroes of the $\beta$-function provide information of the fixed points, in which the physical quantities have no scale dependence.

The electron charge in QED is a very famous case in which a positive $\beta$-function is obtained. In this case, therefore, we expect its observed value to decrease in the low energy regime. Similarly, it is natural to expect the same behavior for a family of parameters describing LV in the SME. If some of their associated $\beta$-functions turn out to be negative, we should expect to observe (unless their scale dependence is weak enough) a relatively large imprint of LV at low energies, which, at the best of our knowledge, has not taken place. Therefore, we can impose constraints on some of the coefficients contained in Lorentz-violating theories by studying their scale dependence; the validity of such models can be tested with this analysis as well.

Indeed, if the coefficients describing LV turn out to have positive $\beta$-functions, we could say that Lorentz and CPT symmetries are low-energy emergent symmetries present in our universe \cite{col1,col2}. While (if the described scenario is correct) the universe and its spacetime structure might be described by very different physics and other symmetry principles at high enough energies, there could exist an energy scale from which such effects are practically washed out. This can help to explain why Lorentz symmetry seems to be exact.

The purpose of this paper is to provide a method to easily solve the $\beta$-functions in some theories with LV and impose constrains on such models by demanding the condition that their $\beta$-functions must be nonnegative|this condition can be used to test the consistency of such models. Particularly, we will apply this method to Yukawa \cite{ref-ferrero3} and electromagnetic theories \cite{ref-kost3,ref-berr} at one-loop order. The $\beta$-functions for a Yukawa theory with Lorentz and CPT violations were already found in \cite{ref-ferrero3}; however, they were not solved. On the other hand, the one-loop order renormalization of Lorentz-violating electrodynamics in flat spacetime was already studied in \cite{ref-kost3}; nonetheless, irreducible representations were not used to decouple the sets of differential equations, allowing us to evaluate the convenience of our method. In addition, by separating the Lorentz-violating operators into different symmetries, the fixed points of the theories can be studied in each particular subspace. It could also be interesting to apply this method to electroweak interactions \cite{ref-collad-1}, non-Abelian gauge theories \cite{ref-collad-2,ref-collad-3}, or other scalar-fermion interactions \cite{ref-ansel} with LV; however, this will not be done here.

This article is organized as follows: in section \ref{sec:yuk} we briefly describe the Lagrange density for a Yukawa theory with Lorentz and CPT violations as well as the different coefficients that describe such theory. Section \ref{sec:renor} describes the general method of solution used to solve for the running couplings. Particularly, such method is applied to Yukawa theories in section \ref{sec:ren-yuk}; the application of such method in a known model such as electromagnetism is studied in section \ref{sec:elec} for comparison reasons. Finally, a brief summary and some conclusions are presented in section \ref{sec:conc}.

\section{Yukawa interactions with LV}\label{sec:yuk}
As stated in \cite{ref-ferrero3}, a theory of Yukawa interactions with Lorentz and CPT violations can be described by the Lagrange density
\begin{align}\label{yuk.1}
\mathcal L_Y=&\,
\frac{_1}{^2}\partial^\mu\phi\partial_\mu\phi+\frac{_1}{^2}K^{\mu\nu}\partial_\mu\phi\partial_\nu\phi
-\frac{_1}{^2}\mu^2\phi^2-\frac{_1}{^{4!}}\lambda\phi^4+\bar\psi\!\left(i\Gamma^\mu\partial_\mu-M\right)\psi
-\phi\bar\psi G\psi,
\end{align}
where irrelevant total-derivative terms have been ignored and
\begin{subequations}
\begin{align}\label{yuk.2}
\Gamma^\nu&=\gamma^\nu\!+\!c^{\mu\nu}\gamma_\mu\!+\!d^{\mu\nu}\gamma_5\gamma_\mu\!+\!e^{\nu}
\!+i\gamma_5f^\nu\!+\!\frac{_1}{^2}g^{\lambda\mu\nu}\sigma_{\lambda\mu}\,,\\
M&=m+i\gamma_5m'+a^\mu\gamma_\mu+b^\mu\gamma_5\gamma_\mu+\frac{_1}{^2}H^{\mu\nu}\sigma_{\mu\nu}\label{yuk.2a}\,,\\
G&=g+i\gamma_5g'+I^\mu\gamma_\mu+J^\mu\gamma_5\gamma_\mu+\frac{_1}{^2}L^{\mu\nu}\sigma_{\mu\nu}\label{yuk.2b}.
\end{align}
\end{subequations}
When quantum corrections are introduced, the previous operators mix according to their symmetry properties. This fact allows us to introduce a set of groups whose elements obey the same symmetries, which are described in table \ref{tab-yuk-1}. The set of counter-terms and the renormalization of the theory is studied in detail in \cite{ref-ferrero3}.

\begin{table}
\begin{center}
{\renewcommand{\arraystretch}{1.4}
\renewcommand{\tabcolsep}{0.2cm}
\begin{tabular}{|c|c|c|c|c|c|c|c|c|}
\hline
Group & \textrm{Operator} & \textrm{C}  & \textrm{P}  & \textrm{T}
& \textrm{CP}       & \textrm{CT} & \textrm{PT} & \textrm{CPT}\\
\hline
1 & $g,\,m,\,c_{00},\,c_{ij},\,K_{00},\,K_{ij}$
& $+$ & $+$ & $+$ & $+$ & $+$ & $+$ & $+$\\
\cline{2-9}
& $g',\,m',\,c_{0j},\,c_{j0},\,K_{0j},\,K_{j0}$
& $+$ & $-$ & $-$ & $-$ & $-$ & $+$ & $+$\\
\hline
2 & $a_0,\,I_0,\,e_0,\,f_j$
& $-$ & $+$ & $+$ & $-$ & $-$ & $+$ & $-$\\
\cline{2-9}
& $a_{j},\,I_j,\,e_j,\,f_{0}$
& $-$ & $-$ & $-$ & $+$ & $+$ & $+$ & $-$\\
\hline
3 & $H_{0j},\,L_{0j},\,d_{00},\,d_{ij}$
& $-$ & $-$ & $+$ & $+$ & $-$ & $-$ & $+$\\
\cline{2-9}
& $H_{ij},\,L_{ij},\,d_{j0},\,d_{0j}$
& $-$ & $+$ & $-$ & $-$ & $+$ & $-$ & $+$\\
\hline
4 & $b_0,\,J_0,\,g_{i00},\,g_{ijk}$
& $+$ & $-$ & $+$ & $-$ & $+$ & $-$ & $-$\\
\cline{2-9}
 & $b_j,\,J_j,\,g_{i0k},\,g_{ij0}$
& $+$ & $+$ & $-$ & $+$ & $-$ & $-$ & $-$\\
\hline
\end{tabular}}
\caption{Discrete symmetry properties that describe the mixing of operators.}
\label{tab-yuk-1}
\end{center}
\end{table}

\section{Renormalization}\label{sec:renor}

In our previous paper \cite{ref-ferrero3}, we found expressions for the $\beta$-functions associated with all the parameters that describe LV in a Yukawa theory. Nonetheless, solving for the running couplings was not performed due to the complexity of the $\beta$-functions. Such difficulty arises when different parameters mix due to quantum corrections and so generating sets of coupled differential equations.

In our analysis, each group of parameters will in turn be decomposed into subgroups using the symmetric group; the action if this group is the permutation of their indices. For example, the coefficient $d^{\,\mu\nu}$ can be decomposed into three irreducible representations: a trace-like part written as $\eta^{\mu\nu}d^{\,\alpha}_{\,\,\,\alpha}$, an antisymmetric part described by $d_A^{\,\mu\nu}=\frac{1}{2}(d^{\,\mu\nu}-d^{\,\nu\mu})$, and a traceless symmetric part given by $d_S^{\,\mu\nu}=\frac{1}{2}(d^{\,\mu\nu}+d^{\,\nu\mu})-\frac{1}{4}\eta^{\mu\nu}d^{\,\alpha}_{\,\,\,\alpha}$.

It is well know that, in this case, the trace-like representation has scalar (rank-zero tensor) properties; the antisymmetric and symmetric parts have vector (rank-one tensor) and rank-two tensor properties respectively. This situation also arises, for instance, when we add the angular momentum of two spin-one particles.

If the mixing of the different coefficients in a Lorentz-violating theory did not exist, we would obtain $\beta$-functions of the form
\begin{align}\label{beta-def}
(\beta_{x})^{\mu_1\cdots\mu_n}=f_{x}(m,m',g,g')x^{\mu_1\cdots\mu_n}\,,
\end{align}
where $f_{x}(m,m',g,g')$ is a function that depends on the masses and the couplings of the Lorentz invariant theory (notice that the sign of $(\beta_{x})^{\mu_1\cdots\mu_n}$ depends on the sign of $f_{x}$). We use the same notation as in \cite{ref-ferrero3,ref-peskin} by defining the dimensionless quantity $\tilde p=p/M$, with $M$ a particular mass scale, to represent the scale associated with the renormalization group. The variable $x^{\mu_1\dots\mu_n}$ or $x_i^{\mu_1\dots\mu_n}(\tilde p)$ will indicate the value of such parameter at the scale $\tilde p$ and $\bar x^{\mu_1\cdots\mu_n}$ its value at $M$ ($\tilde p=1$). For parameters satisfying Eq. (\ref{beta-def}), its scale dependence is given by \cite{ref-ferrero3}
\begin{align}\label{solution}
x^{\mu_1\cdots\mu_n}=\bar x^{\mu_1\cdots\mu_n}\exp\!\left[\int_1^{\tilde p}\frac{d\tilde p\,'}{\tilde p\,'}f_{x}(m,m',g,g')(\tilde p\,')\right]\!.
\end{align}
Nonetheless, when we attempt to obtain the scale dependence of the Lorentz-violating coefficients belonging to the same group, we will obtain coupled systems of differential equations of the form
\begin{align}\label{solution1}
(\beta_{x})^{\mu_1\cdots\mu_n}_m=\sum_{k=1}^{N}M_{mk}x_k^{\mu_1\cdots\mu_n}\,,
\end{align}
where  $M_{mk}$ are matrix elements that depend on some functions $(f_{x})_{mk}$ and $N$ is the number of elements belonging to the same group or subgroup. Each subgroup is in fact described by means of a decomposition into irreducible representations, where all elements belong to the same irreducible class. The symmetric group $S_n$ will be used to accomplish this goal.

If we introduce the diagonal matrix $M^{(d)}$ given by $M=A M^{(d)} A^{-1}$ we can decouple Eq. (\ref{solution1}) and obtain solutions of the form
\begin{align}\label{solution2}
x_k^{(d)\mu_1\cdots\mu_n}&\equiv \bar x_k^{(d)\mu_1\cdots\mu_n} S_{kk}\,,\,\,\,\,\,\textrm{where}\\
S_{kk}&=\exp\!\left[\int_1^{\tilde p}\frac{d\tilde p\,'}{\tilde p\,'}\lambda^{x}_k(m,m',g,g')(\tilde p\,')\right]
\end{align}
are the solutions to the $\beta$-functions in the diagonal basis. Here $\lambda^{x}_k$ are the eigenvalues of the matrix $M$; the initial conditions in the diagonal basis are related to those in the original basis by $\mathbf{\bar x}^{(d)\mu_1\cdots\mu_n}=A^{-1}\mathbf{\bar x}^{\mu_1\cdots\mu_n}$. Therefore, the solutions, in the original basis, are given by
\begin{align}\label{solution2a}
\mathbf{x}^{\mu_1\cdots\mu_n}(\tilde p)&=ASA^{-1}\mathbf{\bar x}^{\mu_1\cdots\mu_n}\,.
\end{align}
Notice that the set of eigenvalues $\lambda_k^{x}$ determine the sign of the $\beta$-functions in the diagonal basis. However, when we have both positive and negative eigenvalues, the coefficients in the original basis have contributions that grow and decrease with the energy scale. This still predicts a large imprint of LV at low-energy scales unless some very special conditions are satisfied or the scale dependence is weak enough. In the presence of fixed points we have null eigenvalues, which indicates that the $\beta$-function of at least one parameter can be written as the linear combination of the other ones.

It should be emphasized that a decomposition into irreducible representations is not the only method that can be used to decouple the sets of differential equations. However, this method greatly reduces the complexity of the problem. For example, let us suppose that we have the set of $\beta$-functions
\begin{subequations}
\begin{align}
(\beta_U)^{\mu\nu}=&\,\frac{_1}{^4}u_1\eta^{\mu\nu}U^{\alpha}_{\,\,\,\alpha}
+\frac{_1}{^4}v_1\eta^{\mu\nu}V^{\alpha}_{\,\,\,\,\alpha}+u_{11}U^{\mu\nu}+u_{12}U^{\nu\mu}+v_{11}V^{\mu\nu}+v_{12}V^{\nu\mu}\,,\\
(\beta_V)^{\mu\nu}=&\,\frac{_1}{^4}u_2\eta^{\mu\nu}U^{\alpha}_{\,\,\,\alpha}
+\frac{_1}{^4}v_2\eta^{\mu\nu}V^{\alpha}_{\,\,\,\,\alpha}+u_{21}U^{\mu\nu}+u_{22}U^{\nu\mu}+v_{21}V^{\mu\nu}+v_{22}V^{\nu\mu}\,,
\end{align}
\end{subequations}
where $u_i$, $u_{ij}$, $v_i$, and $v_{ij}$ are functions of $\tilde p$. If we permute the indices $\{\mu,\nu\}$ we obtain the linear system $\vec\beta_x=M\vec x$, where $\vec x^{\,T}=(U^{\mu\nu},U^{\nu\mu},V^{\mu\nu},V^{\nu\mu},\eta^{\mu\nu}U^{\alpha}_{\,\,\,\alpha},\eta^{\mu\nu}V^{\alpha}_{\,\,\,\alpha})$ and
\begin{align}
M\!=\!\left(\!\!\begin{array}{cccccc}
u_{11} \!&\! u_{12} \!&\! v_{11} \!&\! v_{12} \!&\! \frac{1}{4}u_1            \!&\! \frac{1}{4}v_1 \\
u_{12} \!&\! u_{11} \!&\! v_{12} \!&\! v_{11} \!&\! \frac{1}{4}u_1            \!&\! \frac{1}{4}v_1 \\
u_{21} \!&\! u_{22} \!&\! v_{21} \!&\! v_{22} \!&\! \frac{1}{4}u_2            \!&\! \frac{1}{4}v_2 \\
u_{22} \!&\! u_{21} \!&\! v_{22} \!&\! v_{21} \!&\! \frac{1}{4}u_2            \!&\! \frac{1}{4}v_2 \\
0      \!&\! 0      \!&\! 0      \!&\! 0      \!&\! u_1\!+\!u_{11}\!+\!u_{12} \!&\! y_1\!+\!y_{11}\!+\!y_{12} \\
0      \!&\! 0      \!&\! 0      \!&\! 0      \!&\! u_2\!+\!u_{21}\!+\!u_{22} \!&\! y_2\!+\!y_{21}\!+\!y_{22} \\
\end{array}\!\!\right)\!.
\end{align}
The solution to this problem implies the diagonalization of a $6\times6$ matrix, which is a hard problem. Nevertheless, if we decompose $U^{\mu\nu}$ and $V^{\mu\nu}$ into irreducible representations using Eq. (\ref{eq-g1}) and so the relation $X^{\mu\nu}=\frac{1}{4}\eta^{\mu\nu}X^{\alpha}_{\,\,\,\alpha}+X_A^{\mu\nu}+X_S^{\mu\nu}$ we would obtain the three linear systems
\begin{subequations}
\begin{align}
\left(\!\!\!\begin{array}{c}
(\beta_U)^{\alpha}_{\,\,\,\alpha} \\ (\beta_V)^{\alpha}_{\,\,\,\alpha}
\end{array}\!\!\!\right)
&=\left(\!\!\!\begin{array}{cc}
u_{11}\!+\!u_{12}\!+\!u_1 \!&\! v_{11}\!+\!v_{12}\!+\!v_1\\
u_{21}\!+\!u_{22}\!+\!u_2 \!&\! v_{21}\!+\!v_{22}\!+\!v_2
\end{array}\!\!\!\right)\!\!\!
\left(\!\!\begin{array}{c}
U^{\alpha}_{\,\,\,\alpha} \\ V^{\alpha}_{\,\,\,\alpha}
\end{array}\!\!\!\right)\!,\\
%%%%%%%%%%%%%%%%%%%%%%%%%%%%%%%%%%%%%
\left(\!\!\!\begin{array}{c}
(\beta_U)_A^{\mu\nu} \\ (\beta_V)_A^{\mu\nu}
\end{array}\!\!\!\right)
&=\left(\!\!\begin{array}{cc}
u_{11}-u_{12} & v_{11}-v_{12}\\
u_{21}-u_{22} & v_{21}-v_{22}
\end{array}\!\!\right)\!\!\!
\left(\!\!\begin{array}{c}
U_A^{\mu\nu} \\ V_A^{\mu\nu}
\end{array}\!\!\right)\!,\\
%%%%%%%%%%%%%%%%%%%%%%%%%%%%%%%%%%%%
\left(\!\!\!\begin{array}{c}
(\beta_U)_S^{\mu\nu} \\ (\beta_V)_S^{\mu\nu}
\end{array}\!\!\!\right)
&=\left(\!\!\begin{array}{cc}
u_{11}+u_{12} & v_{11}+v_{12}\\
u_{21}+u_{22} & v_{21}+v_{22}
\end{array}\!\!\right)\!\!\!
\left(\!\!\begin{array}{c}
U_S^{\mu\nu} \\ V_S^{\mu\nu}
\end{array}\!\!\right)\!,
\end{align}
\end{subequations}
which are much easier to solve. The convenience of this procedure is more noticeable when we analyze the $\beta$-functions of higher-rank tensor fields.
The notation introduced in this section will we used in the following ones.

\section{Solutions to the $\beta$-functions of a Yukawa theory}\label{sec:ren-yuk}

\subsection{Lorentz invariant couplings}

Since the $\beta$-functions of the Lorentz-violating parameters depend on the parameters associated with the Lorentz invariant theory, we should find the scale dependence of the latter first. The $\beta$-functions associated with Lorentz-invariant coefficients do not have sign restrictions because their observed values do not vanish at low energy scales. As indicated in \cite{ref-ferrero3}, neglecting $O(c^{\,\alpha}_{\,\,\,\alpha}\,,K^{\alpha}_{\,\,\,\alpha})$ contributions, the couplings $g$ and $g'$ are given by
\begin{eqnarray}\label{eq-run.4}
g(\tilde p)=\bar g F(\tilde p)^{-1/2}\,,\,\,\,\,g'(\tilde p)=\bar g'F(\tilde p)^{-1/2}\,,
\end{eqnarray}
with
\begin{eqnarray}\label{eq-run.3}
F(\tilde p)\,\equiv\,1-5\eta \bar g_+^2\ln\tilde p^{\,2}\,.
\end{eqnarray}
We will use the definitions $\eta^{-1}\equiv16\pi^2$ and $\bar g_\pm^{2}\equiv \bar g^2\pm\bar g'^{\,2}$; the $\tilde p\,$-dependence of $F$ will usually be dropped.
In this theory, the values of both $g$ and $g'$ grow as the energy scale increases and present the usual Landau pole as $F=0$. Fortunately, the Landau pole takes place as $\tilde p_L=\exp\left[\frac{8\pi^2}{5\bar g_+^2}\right]$, which is a very large momentum scale. For $\bar g=\bar g'=\frac{1}{4}$, for instance, $\tilde p_L\sim 10^{54}$, so the perturbative regime includes a very comprehensive energy spectrum.

Having found the scale dependence of $g$ and $g'$, we can now solve for $m$ and $m'$. In terms of the eigenvalues $\lambda_{\pm}=\frac{\eta}{2}(\,g_+^2\pm\sqrt{(5g^2+3g'^{\,2})^2+4g^2g'^{\,2}}\,)$ we obtain
\begin{equation}
m_\pm^{(d)}=\bar m^{(d)}_\pm F^{-\frac{8\pi^2}{5\bar g_+^2}\bar\lambda_{\pm}}
\equiv \bar m_\pm^{(d)}F_{\pm}\,.
\end{equation}
Hence, in terms of $F^{(\pm)}\equiv F_+\pm F_-$
\begin{subequations}
\begin{align}\label{eq:m1}
m&=\frac{_1}{^2}\bar mF^{(+)}+
\frac{\bar m(5\bar g^2-3\bar g'^{\,2})+8\bar m'\bar g\bar g'}{4\eta^{-1}\bar\lambda_+-2\bar g_+^2}F^{(-)},
\\ \label{eq:m2}
m'&=\frac{_1}{^2}\bar m'F^{(+)}
-\frac{\bar m'(5\bar g^2-3\bar g'^{\,2})-8\bar m\bar g\bar g'}{4\eta^{-1}\bar\lambda_+-2\bar g_+^2}F^{(-)}.
\end{align}
\end{subequations}
\begin{figure}
\begin{minipage}[t]{0.45\linewidth}
\centering
\includegraphics[scale=0.67]{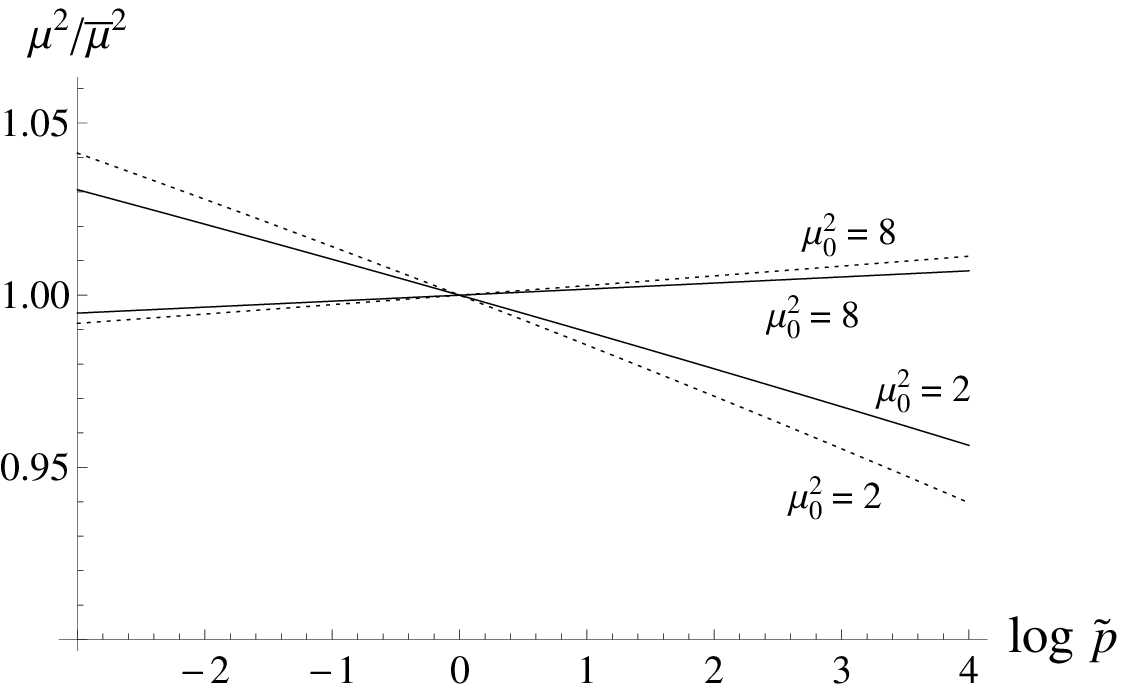}
\caption{Boson mass as a function of $\tilde p$. Here $\mu_0^2\equiv\frac{\bar \mu^2}{m_0^2}$. In the solid lines we used the conditions $\bar m=\bar m'=m_0$, $\bar g=0.2$, and $\bar g'=0.1$. For the dashed lines $\bar m=\bar m'=m_0$, $\bar g=0.1$, and $\bar g'=0.25$. $m_0$ is an arbitrary mass scale.}
\label{fig:1}
\end{minipage}
\hspace{0.5cm}
\begin{minipage}[t]{0.45\linewidth}
\centering
\includegraphics[scale=0.71]{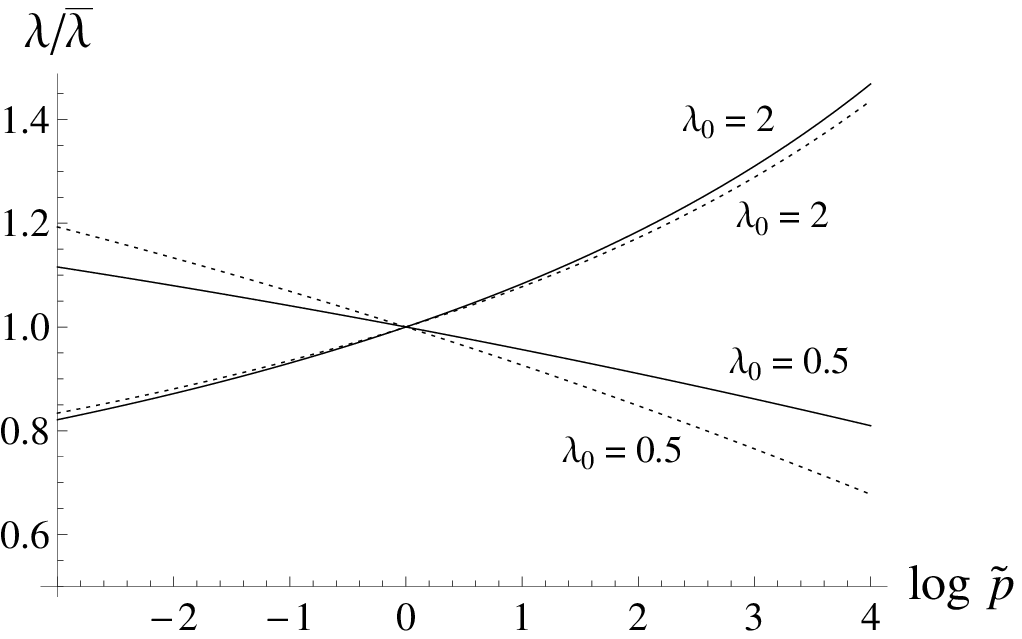}
\caption{Scalar coupling as a function of $\tilde p$. Here $\lambda_0\equiv\frac{\bar \lambda}{g_0}$. In the solid lines we used the conditions $\bar g=0.2\bar g_0$ and $\bar g'=0.1g_0$. For the dashed lines $\bar g=0.1g_0$ and $\bar g'=0.25g_0$. $g_0$ is an arbitrary constant.}
\label{fig:2}
\end{minipage}
\end{figure}
In order to find the momentum dependence of the boson mass and the scalar coupling $\lambda$, we need to solve, respectively, the differential equations
\begin{eqnarray}
\tilde p\frac{d\mu^2}{d\tilde p}\!\!\!&=&\!\!\!8\eta\Big[\mu^2g_+^2-2(mg+m'g')^2-(m^2+m'^{\,2})g_+^2\Big]\,,\\
\tilde p\,\frac{d\lambda}{d\tilde p}\!\!\!&=&\!\!\!\eta\Big[3\lambda^2+8\lambda g_+^2-48g_+^2\Big]\,,
\end{eqnarray}
where $O(c^{\,\mu}_{\,\,\,\mu})$ and $O(K^{\mu}_{\,\,\,\mu})$ have been neglected.
%From last equations we see that $\mu^2$ and $\lambda$ increase with the energy scale if $\mu^2>2\frac{(mg+m'g')^2}{g_+^2}+m^2+m'^{\,2}$ and
%$3\lambda^2+8\lambda g_+^2-48 g_+^2>0$ respectively.
Since we have not been able to obtain analytical solutions for the last set of equations, some numerical solutions are shown in figures \ref{fig:1} and \ref{fig:2}.

\subsection{Lorentz-violating parameters}

Now we will solve for the running couplings associated with the Lorentz-violating theory.
We will independently focus on each one of the four groups presented in table \ref{tab-yuk-1}.

\subsection{Solution to group 1}
For this family we have the set of differential equations \cite{ref-ferrero3}
\begin{subequations}
\begin{align}
(\beta_K)^{\mu\nu}&=4\eta g_+^2\bigl(K^{\mu\nu}+c^{\,\alpha}_{\,\,\,\alpha}\eta^{\mu\nu}-c^{\,\mu\nu}-c^{\,\nu\mu}\bigr),\\
(\beta_c)^{\,\mu\nu}&=\frac{_1}{^3}\eta g_+^2\bigl(c^{\,\mu\nu}\!+c^{\,\nu\mu}-K^{\mu\nu}\!
+(c^{\,\alpha}_{\,\,\,\alpha}\!+\!K^{\,\alpha}_{\,\,\,\alpha})\eta^{\mu\nu}\bigr).
\end{align}
\end{subequations}
In order to solve previous equations we will use the rank-two irreducible decomposition
\begin{align}\label{eq-g1}
X^{\mu\nu}_A&=\frac{_1}{^2}(X^{\mu\nu}\!-\!X^{\nu\mu}),\,\,\,\,
X_S^{\mu\nu}=\frac{_1}{^2}(X^{\mu\nu}\!+\!X^{\nu\mu})
\!-\frac{_1}{^4}\eta^{\mu\nu}X^\alpha_{\,\,\alpha}.
\end{align}
Using this decomposition we find that $(\beta_K)_A^{\mu\nu}=(\beta_c)_A^{\mu\nu}=0$; this is expected in the case of $K_{\mu\nu}$ because it is symmetric by definition. Writing our linear system as in Eq. (\ref{solution1}) with
$(\vec x^{\,T})^{\mu\nu}=(K_S^{\mu\nu},c_S^{\mu\nu})$ (and something similar for the traces) we find the solutions
\begin{subequations}
\begin{align}
K^{(d)\alpha}_{\phantom{mm}\alpha}&=\bar K^{(d)\alpha}_{\phantom{mm}\alpha}F^{-3/5}\,,\,\,\,\,\,
c^{(d)\alpha}_{\phantom{mm}\alpha}=\bar c^{(d)\alpha}_{\phantom{mm}\alpha}\,,\\
K^{(d)\mu\nu}_S&=\bar K^{(d)\mu\nu}_S F^{-7/15}\,,\,\,\,\,\,
c^{(d)\mu\nu}_S=\bar c^{(d)\mu\nu}_S.
\end{align}
\end{subequations}
So, in the original basis
\begin{subequations}
\begin{align}\label{g3.ab}
K^{\,\alpha}_{\,\,\,\alpha}&=-\frac{_4}{^3}\bar c^{\,\alpha}_{\,\,\,\alpha}\bigl(1-F^{-3/5}\bigr)
+\frac{_1}{^3}\bar K^{\,\alpha}_{\,\,\,\alpha}\bigl(1+2F^{-3/5}\bigr),\\
%%%%%%%%%%%%%%%%%%%%%%%%%%%%%%%%%
c^{\,\alpha}_{\,\,\,\alpha}&=\frac{_1}{^3}\bar c^{\,\alpha}_{\,\,\,\alpha}\bigl(2+F^{-3/5}\bigr)
-\frac{_1}{^6}\bar K^{\,\alpha}_{\,\,\,\alpha}\bigl(1-F^{-3/5}\bigr),\\
%%%%%%%%%%%%%%%%%%%%%%%%%%%%%%%%%%
K^{\mu\nu}_{S}&=\frac{_{12}}{^7}\bar c^{\,\mu\nu}_{S}\bigl(1-F^{-7/15}\bigr)
+\frac{_1}{^7}\bar K^{\mu\nu}_{S}\bigl(1+6F^{-7/15}\bigr),\\
%%%%%%%%%%%%%%%%%%%%%%%%%%%%%%%%%
c^{\,\mu\nu}_{S}&=\frac{_1}{^7}\bar c^{\,\mu\nu}_{S}\bigl(6+F^{-7/15}\bigr)
+\frac{_1}{^{14}}\bar K^{\mu\nu}_{S}\bigl(1-F^{-7/15}\bigr).
\end{align}
\end{subequations}

\subsection{Group 2}

Having solved for the coefficients $c^{\,\mu\nu}$ and $K^{\mu\nu}$, we proceed with the vector parameters, which are $a^\mu, e^\mu, f^{\mu}$ and $I^\mu$. Since all of them are rank-one tensors, they do not need to be decomposed into irreducible representations. In this case, however, the $\beta$-functions of the linear system do not explicitly depend on $a^\mu$. Consequently, we must exclude $a^\mu$ from the linear system, which is not invertible, and find its scale dependence after determining the other three solutions. Thus we must solve the set of differential equations \cite{ref-ferrero3}
\begin{eqnarray}\label{eq-g11}
\left(\!\!\!\begin{array}{c}(\beta_I)^\mu \\ (\beta_f)^\mu \\ (\beta_e)^\mu\end{array}\!\!\!\right)
\!\!\!\!&=&\!\!\!\eta\left(\!\!\!\begin{array}{ccc}
6g_+^2 & -g'g_+^2 & -gg_+^2\\
-2g' & 2g'^{\,2} & 2gg'  \\
-2g & 2gg'& 2g^2 \\
\end{array}\!\!\!\right)\!\!\!
\left(\!\!\!\begin{array}{c}I^\mu \\ f^\mu \\ e^\mu\end{array}\!\!\!\right)\!,
\end{eqnarray}
which generate the solutions
%\begin{subequations}
\begin{align}
&I^{(d)\mu}=\bar I^{(d)\mu}\,,\,\,\,
f^{(d)\mu}=\bar f^{(d)\mu}F_+\,,\,\,\,
e^{(d)\mu}=\bar e^{(d)\mu}F_-\,,\nonumber\\
&\textrm{with}\,\,\,\,\,F_{\pm}(\tilde p)\equiv F^{-(4\mp\sqrt{6})/10}\,.
\end{align}
%\end{subequations}
In the original basis we have, with $F^{(\pm)}\equiv F_+\pm F_-$
\begin{subequations}
\begin{align}\label{eq-g12}
I^\mu&=\frac{_1}{^2}\bar I^\mu F^{(+)}\!+\!
\frac{_1}{^{2\sqrt{6}}}\bigl(\bar g\bar e^\mu+\bar g'\bar f^\mu-2\bar I^\mu\bigr)F^{(-)},\\
%%%%%%%%%%%%%%%%%%%%%%%%%%
f^\mu&=\frac{\bar g}{\bar g_+^2}\bigl(\bar g\bar f^\mu-\bar g'\bar e^\mu\bigr)
+\frac{\bar g'\bar I^\mu}{\sqrt{6}\,\bar g_+^2}F^{(-)}
+\frac{\bar g'(\bar g\bar e^\mu+\bar g'\bar f^\mu)}{2\sqrt{6}\,\bar g_+^2}
\bigl(2F^{(-)}+\sqrt{6}F^{(+)}\bigr),\\ \label{eq-g12b}
%%%%%%%%%%%%%%%%%%%%%%%%%
e^\mu&=\frac{\bar g'}{\bar g_+^2}\bigl(\bar g'\bar e^\mu-\bar g\bar f^\mu\bigr)
+\frac{\bar g\bar I^\mu}{\sqrt{6}\,\bar g_+^2}F^{(-)}+\frac{\bar g(\bar g\bar e^\mu+\bar g'\bar f^\mu)}
{2\sqrt{6}\,\bar g_+^2}\bigl(2F^{(-)}+\sqrt{6}F^{(+)}\bigr).
\end{align}
\end{subequations}
On the other hand, $a^{\,\mu}$ can be determined by solving \cite{ref-ferrero3}
\begin{align}\label{eq-g13}
a^{\,\mu}=\bar a^{\,\mu}+\eta\!\int_1^{\tilde p}\!\frac{d\tilde p\,'}{\tilde p\,'}
\Big[4(gm\!+\!g'm')I^\mu\!-g_+^2(me^\mu\!+\!m'\!f^\mu)\!\Big]
\end{align}
using Eqs. (\ref{eq-run.4}), (\ref{eq:m1}), (\ref{eq:m2}), and (\ref{eq-g12})$-$(\ref{eq-g12b}). Although $a^\mu$ can be solved analytically, its expression is quite long and so it will not be shown here. An interesting fact though is to study its leading contribution at low momentum ($\tilde p\ll1$); it is given by
\begin{eqnarray}\label{eq-g13a}
a^{\,\mu}(\tilde p)\sim\bar a^{\,\mu}+A\bar I^\mu\bigl(1-F^{\alpha}\bigr),
\end{eqnarray}
where $\alpha=\frac{1}{20}\left(1+2\sqrt{6}+\frac{1}{\bar g_+^2}\sqrt{(5\bar g^2+3\bar g'^{\,2})^2+4\bar g^2\bar g'^{\,2}}\right)$ and $A$ is a constant with units of mass. The maximum value of $\alpha$ is achieved as $g'=0$, thus $\alpha_{\textrm{max}}=\frac{1}{10}(3+\sqrt{6}\,)$. In spite that $\alpha>0$ and so $a^{\mu}$ increases as $\tilde p\to 0$, we do not really expect to observe large values of $a^\mu$ at low energy scales. For $\bar g=0.25$, $\bar g'=0$, and $\tilde p=10^{-100}$, for example, $F^{\alpha_{\textrm{max}}}\simeq1.42$. So $a^\mu =O(\bar a^\mu,A\bar I^\mu)$ at such low scales.

\subsection{Group 3}

In order to solve for $d^{\,\mu\nu},\,H^{\mu\nu}$ and $L^{\mu\nu}$ ($H^{\mu\nu}$ and $L^{\mu\nu}$ are antisymmetric
by construction), we notice that they obey the set of differential equations \cite{ref-ferrero3}
\begin{subequations}
\begin{align}\label{eq-g14}
(\beta_d)^{\mu\nu}&=\eta\Big[\frac{_1}{^3}g_+^2(5d^{\,\mu\nu}\!-d^{\,\nu\mu}\!-\eta^{\mu\nu}d^{\,\alpha}_{\,\,\,\alpha})
\!-2g'L^{\mu\nu}_A\!-\!2gL_A^{*\mu\nu}\Big],
\\\label{eq-g14a}
%%%%%%%%%%%%%%%%%%%%%%%%%%%%%%%%%
(\beta_H)^{\mu\nu}_A&=\eta\Big[g_+^2H^{\mu\nu}_A\!+4(gm+g'm')L^{\mu\nu}_A\!+\!4(mg'-m'g)L_A^{*\mu\nu}
\nonumber\\
%%%%%%%
&\phantom{00}+2(mg_-^2+2m'gg')d_A^{*\mu\nu}\!+2(m'g_-^2-2mgg')d_A^{\,\mu\nu}\Big],
\\\label{eq-g14b}
%%%%%%%%%%%%%%%%%%%%%%%%%%%%%%%%
(\beta_L)_A^{\mu\nu}&=\eta g_+^2\Big[7L^{\mu\nu}_A+2gd_A^{*\mu\nu}-2g'd^{\,\mu\nu}_A\Big],
\end{align}
\end{subequations}
where we used the Hodge dual representation $X^{*\mu\nu}=\frac{1}{2}\varepsilon^{\mu\nu\alpha\beta}X_{\alpha\beta}$. (From now on we will use the convention $\varepsilon^{0123}=-\varepsilon_{0123}=1$.)
After decomposing $d^{\,\mu\nu}$ into its irreducible parts we find
\begin{subequations}
\begin{align}\label{eq-g15}
d^{\,\alpha}_{\,\,\,\alpha}&=\bar d^{\,\alpha}_{\,\,\,\alpha}\,,\,\,\,\,\,
d_S^{\,\mu\nu}=\bar d_S^{\,\mu\nu}F^{-2/15}.
\end{align}
\end{subequations}
The antisymmetric part of $d^{\,\mu\nu}$ couples with Eqs. (\ref{eq-g14a}) and (\ref{eq-g14b}) and satisfies
\begin{subequations}
\begin{align}\label{eq-g16}
(\beta_d)^{\mu\nu}_A&=2\eta\bigl(g_+^2d_A^{\,\mu\nu}-g'L^{\mu\nu}_A-gL_A^{*\mu\nu}\bigr).
\end{align}
\end{subequations}
In spite that Eqs. (\ref{eq-g14a}), (\ref{eq-g14b}), and (\ref{eq-g16}) only involve antisymmetric representations, the Hodge dual components are also present. Such system cannot be solved directly because, although the Hodge dual representation of an antisymmetric tensor carries the same information and should have the same scale behavior than the original tensor itself, different components are mixed. For example, for $\mu=0$ and $\nu=1$, $L^{\mu\nu}_A=L_A^{01}$ and $L^{*\mu\nu}_A=L_A^{23}$. This would mix different components of $L^{\mu\nu}_A$, for instance, in Eq. (\ref{eq-g16}).

The solution to this issue is solved by taking the Hodge dual of Eqs. (\ref{eq-g14a}), (\ref{eq-g14b}), and (\ref{eq-g16}) to obtain a six-variables linear system. In this case we can define the six-component vector $(\vec x^{\,T})^{\mu\nu}=(d_A,d_A^*,H_A,H_A^*,L_A,L_A^*)^{\mu\nu}$, where, for example, we have the condition $(\beta_d)^{*\mu\nu}_A=2\eta\bigl(g_+^2d_A^{*\mu\nu}-g'L^{*\mu\nu}_A+gL_A^{\mu\nu}\bigr)$  from Eq. (\ref{eq-g16}). We thus find the solutions
\begin{subequations}
\begin{align}\label{eq-g21a}
d_A^{\,(d)\mu\nu}&=\bar d_A^{\,(d)\mu\nu}F_1\,,\,\,\,\,
F_1(\tilde p)\equiv F^{-1/10}\,,\\
%%%%%%%%%%%%%%%%%%%
H_A^{\,(d)\mu\nu}&=\bar H_A^{\,(d)\mu\nu}F_2\,,\,\,\,\,
F_2(\tilde p)\equiv F^{-\frac{9-\sqrt{2}}{20}}\,,\\
%%%%%%%%%%%%%%%%%%%
L_A^{\,(d)\mu\nu}&=\bar L_A^{\,(d)\mu\nu}F_3\,,\,\,\,\,
F_3(\tilde p)\equiv F^{-\frac{9+\sqrt{2}}{20}}\,.
\end{align}
\end{subequations}
Using previous definitions and $F_{23}^{(\pm)}\equiv F_2\pm F_3$ we find
\begin{subequations}
\begin{align}\label{eq-g22}
d_A^{\,\mu\nu}&=\frac{_1}{^2}\bar d_A^{\,\mu\nu}\!\Big[F_{23}^{(+)}\!+\frac{_{5F_{23}^{(-)}}}{^{\sqrt{41}}}\Big]
\!+\frac{_{2F_{23}^{(-)}}}{^{\sqrt{41}\,\bar g_+^2}}\bigl(\bar g\bar L_A^{*\mu\nu}\!+\bar g'\bar L_A^{\mu\nu}\bigr),\\
H_A^{\mu\nu}&=\bar H_A^{\mu\nu}F_1
+\frac{_1}{^{\bar g_+^2}}\Big[\bigl(2\bar m\bar g\bar g'\!-\bar m'\bar g_-^2\bigr)\bar d_A^{\,\mu\nu}
\!-\!\bigl(2\bar m'\bar g\bar g'\!+\!\bar m\bar g_-^2\bigr)\bar d_A^{*\mu\nu}\Big]\Big[2F_1-F_{23}^{(+)}-\frac{_{5F_{23}^{(-)}}}{^{\sqrt{41}}}\Big]
\nonumber\\
&\phantom{0i}-\frac{_{4F_{23}^{(-)}}}{^{\sqrt{41}\,\bar g_+^2}}\Big[(\bar m\bar g+\bar m'\bar g')\bar L_A^{\mu\nu}
+(\bar m\bar g'\!-\bar m'\bar g )\bar L_A^{*\mu\nu}\Big],\\
L_A^{\mu\nu}&=\frac{_1}{^2}\bar L_A^{\mu\nu}\!\Big[F_{23}^{(+)}\!-\frac{_{5F_{23}^{(-)}}}{^{\sqrt{41}}}\Big]
\!+\frac{_{2F_{23}^{(-)}}}{^{\sqrt{41}}}\bigl(\bar g'\bar d_A^{\,\mu\nu}-\bar g\bar d_A^{\,*\mu\nu}\bigr).
\end{align}
\end{subequations}

\subsection{Group 4}

This set of differential equations is the most difficult to handle; it couples the terms $g^{\lambda\mu\nu},\,J^\mu$, and $b^{\,\mu}$, as well as their dual representations. The difficulty mainly arises on finding an appropriate irreducible decomposition for $g^{\lambda\mu\nu}$; a careful decomposition is performed in appendix \ref{appAa}. An anti-symmetrization in the indices $\{\lambda,\mu\}$ of $g^{\lambda\mu\nu}$ must be performed so it can be written according to Eqs. (\ref{gestil1})$-$(\ref{gestil6}). 

Using the relation $(\beta_g)^{\lambda\mu\nu}=(\beta_g)^{\lambda\mu\nu}_A+\frac{4}{3}(\beta_g)^{\lambda\mu\nu}_{1^-}
+\frac{1}{3}\eta^{\lambda\nu}(\beta_g)^\mu_2-\frac{1}{3}\eta^{\mu\nu}(\beta_g)^\lambda_2$ and after decomposing into irreducible representations we have,
in terms of $g^{[*\lambda\mu]\nu}\equiv\frac{1}{2}\varepsilon^{\alpha\beta\lambda\mu}g_{\alpha\beta}^{\phantom{mi}\nu}$ \cite{ref-ferrero3}
\begin{align}\label{betages}
(\beta_g)^{\lambda\mu\nu}=&\,
2\eta\Big[g^2g_A^{\lambda\mu\nu}\!+gJ^{*\lambda\mu\nu}\!+\!\frac{_2}{^{9}}gg'g_2^{*\lambda\mu\nu}-\frac{_2}{^9}gg'g_3^{*\lambda\mu\nu}
+\frac{_4}{^9}(2g^2+g'^{\,2})g_{1^-}^{\lambda\mu\nu}-\frac{_4}{^9}g_-^2g_{2^+}^{\lambda\mu\nu}
\nonumber\\
&+\eta^{\lambda[\nu]}\bigl(\frac{_1}{^{18}}(g^2+5g'^{\,2})g_2+\frac{_1}{^9}g_-^2g_1+g'J\bigr)^{\mu}
+gg'g_A^{[*\lambda\mu]\nu}-\frac{_4}{^9}gg'g_{1^-}^{[*\lambda\mu]\nu}+\frac{_8}{^9}gg'g_{2^+}^{[*\lambda\mu]\nu}\Big].
\end{align}
where $X^{\lambda[\nu]\mu}\equiv X^{\lambda\nu\mu}-X^{\mu\nu\lambda}$.
Multiplying last equation by $\varepsilon_{\lambda\mu\nu\sigma}$ we obtain
\begin{align}\label{betages2}
(\beta_g)^{\!*\,\mu}_A&=\eta\Big[2g^2g_A^{*\mu}+2gJ^{\mu}
+\frac{_4}{^9}gg'g_1^{\mu}+\frac{_8}{^9}gg'g_2^{\mu}\Big]\,,
\end{align}
where we used the condition $g_1^\mu+g_2^\mu+g_3^\mu=0$ (see appendix \ref{appAa} for more details).

\begin{figure}[t]
\centering
\includegraphics[scale=0.75]{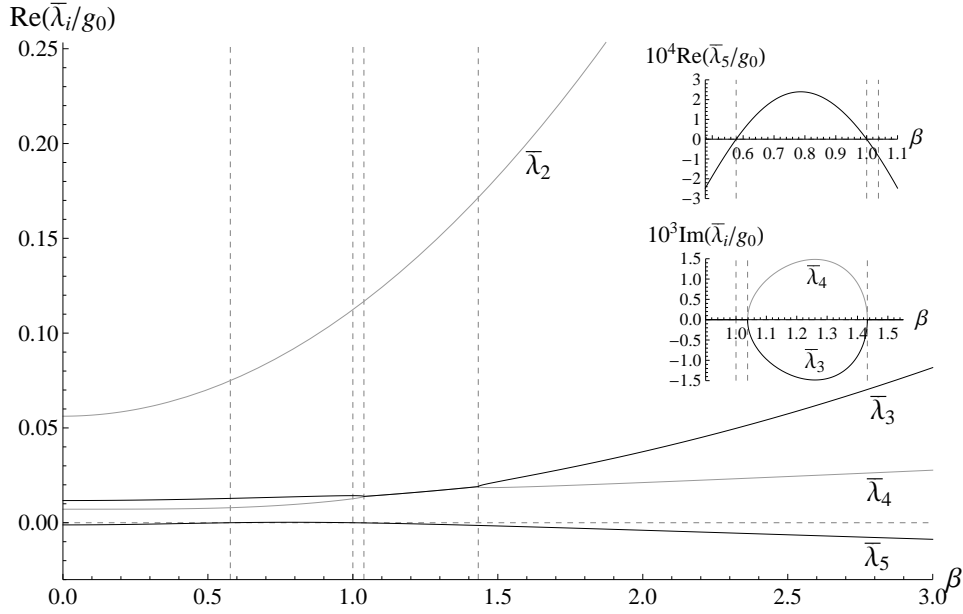}
\caption{Values of the eigenvalues $\bar\lambda_2,\bar\lambda_3,\bar\lambda_4$ and $\bar\lambda_5$ as a function of $\bar g$ and $\bar g'$. Here $\bar g=g_0$ and $\bar g'=\beta g_0$. The vertical dashed lines divide a set of intervals; the horizontal dashed line is the value 0. The first interval is $0\leq\beta<\frac{1}{\sqrt{3}}$ and all eigenvalues except $\bar\lambda_5$ are positive, like in the third interval given by $1<\beta<1.038$ and the fifth, where $\beta\geq1.434$. In the second interval all eigenvalues are positive and $\frac{1}{\sqrt{3}}\leq\beta\leq1$. In the fourth interval, $1.038<\beta<1.433$, $\bar\lambda_3$ and $\bar\lambda_4$ are complex. The upper small graph zooms the interval where $\lambda_5\geq0$, the lower small graph zooms the imaginary parts of $\bar\lambda_3$ and $\bar\lambda_4$.}
\label{fig:3}
\end{figure}

The $\beta$-function of $g_2^\mu$, as defined in Eq. (\ref{traces3}), can be found by multiplying Eq. (\ref{betages}) by $\eta_{\lambda\nu}$. In order to find $(\beta_g)_1^\mu$ we must perform index
permutations in Eq. (\ref{betages}) before contracting the first two indices (this must be done because Eq. (\ref{betages}) does not provide enough information). After doing this we find
\begin{subequations}
\begin{align}\label{betages5}
(\beta_g)_2^\mu&=\eta\Big[\frac{_1}{^{3}}(g^2\!+5g'^{\,2})g_2^\mu+\frac{_2}{^3}g_-^2g_1^\mu+6g'J^\mu\!+6gg'g_A^{*\mu}\Big],\\
%%%%%%%%%%%%%%%%%%%%%%%%%%%%%%%
(\beta_g)_1^\mu&=\eta\Big[\frac{_2}{^{3}}(2g^2\!+g'^{\,2})g_1^\mu+\frac{_7}{^{6}}g_-^2g_2^\mu-3g'J^\mu\!-3gg'g_A^{*\mu}\Big].
\end{align}
\end{subequations}
The $\beta$-functions of $b^{\,\mu}$ and $J^\mu$ were already found in \cite{ref-ferrero3}. Defining 
$(\vec x^{\,T})^\mu\equiv(b,J,g_1,g_2,g_A^*)^{\mu}$ we obtain, according to Eq. (\ref{solution1}), the matrix
\begin{eqnarray}\label{eq-mat0}
M\!=\!\eta\!\left(\!\!\!\begin{array}{ccccc}
g_+^2 \!\!&\!\! 2(mg\!+\!m'g') \!\!\!&\!\!\! 0                              \!\!\!&\!\!\! \frac{1}{2}m'g_+^2             \!\!&\!\! \frac{3}{2}mg_+^2\\
0     \!\!&\!\! 8g_+^{2}       \!\!\!&\!\!\! 0                              \!\!\!&\!\!\! g'g_+^2                        \!\!&\!\! 3gg_+^2\\
0     \!\!&\!\! -3g'           \!\!\!&\!\!\! \frac{2}{3}(2g^2\!+\!g'^{\,2}) \!\!\!&\!\!\! \frac{7}{6}g_-^2               \!\!&\!\! -3gg'\\
0     \!\!&\!\! 6g'            \!\!\!&\!\!\! \frac{2}{3}g_-^2               \!\!\!&\!\!\! \frac{1}{3}(g^2\!+\!5g'^{\,2}) \!\!&\!\! 6gg'\\
0     \!\!&\!\! 2g             \!\!\!&\!\!\! \frac{4}{9}gg'                 \!\!\!&\!\!\! \frac{8}{9}gg'                 \!\!&\!\! 2g^2
\end{array}\!\!\!\right)\!.
\end{eqnarray}
While one of the eigenvalues (the one associated with $b^{\,\mu}$) of the previous matrix is always $\lambda_1=\eta g_+^2$ and so is positive, the other four eigenvalues take very complicated forms and their signs depend on the values of $g$ and $g'$; this behavior is shown in figure \ref{fig:3}.
For some values of such couplings we have a negative eigenvalue and even complex results. The complex results associated with $\lambda_3$ and $\lambda_4$ in figure \ref{fig:3} are, however, just a mathematical artifact and the final results are real in the original basis (this will become more clear below); since their real parts  are always positive, they do not generate an asymptotically-free behavior.

In spite that $\lambda_5$ is negative for most values of $g$ and $g'$, the evolution of the parameters described by the matrix in Eq. (\ref{eq-mat0}) also depends from electromagnetic contributions \cite{ref-kost3}. Hence, the inclusion of electromagnetic effects may change such tendency. Additionally, the magnitude of $\lambda_5$ does not take considerably large values for typical values of $\beta$ (this parameter is defined in figure \ref{fig:3}). In figure \ref{fig:3} the smallest value of $\bar\lambda_5$ is $\bar\lambda_5\simeq-8.8\times 10^{-3}$ and takes place at $\beta=3$. Using this value and $\bar g=0.25$, we see that $F^{-|\bar\lambda_5|}\simeq0.96$ for $\tilde p=10^{-100}$, which is very close to 1. Therefore, $\lambda_5$ induces a negligible deviation at low energy scales.

If we neglect electromagnetic effects, there are still some conditions that could make all eigenvalues in Eq. (\ref{eq-mat0}) positive for arbitrary values of $g$ and $g'$. The solution is assuming that the values of some coefficients at the renormalization scale vanish, so there is no LV at any energy scale coming from such fields. Nevertheless, such conditions are very unlikely and restrictive and demand fine-tuning.

Using Eq. (\ref{eq-mat0}) into Eq. (\ref{betages}) and the $\beta$-function of $g^{\lambda\nu\mu}$ (last one can be found by permuting indices) we have that the other two irreducible representations satisfy
\begin{subequations}
\begin{align}
(\beta_g)^{\lambda\mu\nu}_{1^-}
=&\,\frac{_2}{^3}\eta\Big[(2g^2\!+\!g'^{\,2})g_{1^-}^{\lambda\mu\nu}\!\!-g_-^2g_{2^+}^{\lambda\mu\nu}
-gg'g_{1^-}^{[*\lambda\mu]\nu}\!\!+2gg'g_{2^+}^{[*\lambda\mu]\nu}\Big],\\
%%%%%%%%%%%%%%%%%%%%%%%%%%%%%%%%%%%%%%%%%
(\beta_g)^{\lambda\mu\nu}_{2^+}
=&\,\frac{_2}{^3}\eta\Big[(2g^2\!+\!g'^{\,2})g_{2^+}^{\lambda\mu\nu}\!\!-g_-^2g_{1^+}^{\lambda\mu\nu}
-gg'g_{1^-}^{[*\lambda\mu]\nu}\!\!+2gg'g_{2^+}^{[*\lambda\mu]\nu}\Big].
\end{align}
\end{subequations}
Last system is, however, very difficult to solve. For convenience we will change to a different basis by defining 
$g_{\pm}^{\lambda\mu\nu}\equiv g_{1^-}^{\lambda\mu\nu}\pm \, g_{2^+}^{\lambda\mu\nu}$. In terms of 
$(\vec x^{\,T})^{\lambda\mu\nu}=(g^{\lambda\mu\nu}_{-},g^{[\!*\lambda\mu]\nu}_{-},g^{\lambda\mu\nu}_{+},g^{[\!*\lambda\mu]\nu}_{+})$ and the matrix
\begin{align}
M&=\frac{2\eta}{3}
\left(\!\!\!\begin{array}{cccc}
3g^2 \!\!&\!\! 0     \!\!&\!\! 0             \!\!&\!\! 0 \\
0    \!\!&\!\! 3g^2  \!\!&\!\! 0             \!\!&\!\! 0 \\
0    \!\!&\!\! -3gg' \!\!&\!\! g^2+2g'^{\,2} \!\!&\!\! gg' \\
3gg' \!\!&\!\! 0     \!\!&\!\! -gg'          \!\!&\!\! g^2+2g'^{\,2} \\
\end{array}\!\!\!\!\right)\!
\end{align}
we find the eigenvalues $\lambda_{1,2}=2\eta g^2$ and $\lambda_{3,4}=g^2+2g'^{\,2}\mp igg'$. Although we have two complex eigenvalues, the evolution of the given parameters is real when we go back to the original basis (this also applies for the parameters described in figure \ref{fig:3}).
In the diagonal basis we find
\begin{subequations}
\begin{align}
g_-^{(d)\lambda\mu\nu}&=\bar g_-^{(d)\lambda\mu\nu}F_1\,,\,\,\,\,
g_-^{(d)[*\lambda\mu]\nu}=\bar g_-^{(d)[*\lambda\mu]\nu}F_1^*\,,\\
g_+^{(d)\lambda\mu\nu}&=\bar g_+^{(d)\lambda\mu\nu}F_2\,,\,\,\,\,
g_+^{(d)[*\lambda\mu]\nu}=\bar g_+^{(d)[*\lambda\mu]\nu}F_2^*\,,\\
F_1(\tilde p)&=F^{-\bar g^2\!/5\bar g_+^2}\,,\,\,\,\,\\\label{eq:F2}
F_2(\tilde p)&=F^{-(\bar g^2+2\bar g'^2)/(15\bar g_+^2)}
\exp\!\Big(i\frac{_{\bar g\bar g'\ln F}}{^{15\bar g_+^2}}\Big)\,.
\end{align}
\end{subequations}
In terms of %$g_{\pm}^{\lambda\mu\nu}=g_{1^-}^{\lambda\mu\nu}\pm g_{2^+}^{\lambda\mu\nu}$,
$F_2^R\equiv\textrm{Re}\bigl[F_2\bigr]$ and $F_2^I\equiv\textrm{Im}\bigl[F_2\bigr]$ we have
\begin{subequations}
\begin{align}
g_{1^-}^{\lambda\mu\nu}&=\frac{_1}{^2}\bar g_-^{\lambda\mu\nu}F_1+\frac{_1}{^2}\bar g_+^{\lambda\mu\nu}F_2^R
-\frac{_1}{^2}\bar g_+^{[\!*\lambda\mu]\nu}F_2^I
+\frac{3\bar g^2\bar g'^{\,2}\bar g_-^{[\!*\lambda\mu]\nu}\!\!+6\bar g_-^{2}\bar g\bar g'\bar g_-^{\lambda\mu\nu}}
{8(\bar g_-^2)^2+2\bar g^2\bar g'^{\,2}}F_2^I\nonumber\\
%%%%%%%%%%%%
&\phantom{=}+\frac{3\bar g^2\bar g'^{\,2}\bar g_-^{\lambda\mu\nu}-6\bar g_-^{2}\bar g\bar g'\bar g_-^{[\!*\lambda\mu]\nu}}
{8(\bar g_-^2)^2+2\bar g^2\bar g'^{\,2}}\bigl(F_1-F_2^R\bigr),\\
%%%%%%%%%%%%%%%%%%%%%%%%%%%%%%%%%%%%%%%%%%%
%%%%%%%%%%%%%%%%%%%%%%%%%%%%%%%%%%%%%%%%%%%
g_{2^+}^{\lambda\mu\nu}&= g_{1^-}^{\lambda\mu\nu}-\bar g_-^{\lambda\mu\nu}F_1.
\end{align}
\end{subequations}
At first glance, one might be tempted to think that $g_{1^-}^{\lambda\mu\nu}$ and $g_{2^+}^{\lambda\mu\nu}$ have an oscillatory behavior because of the definition of $F_2$ in Eq. (\ref{eq:F2}). Nevertheless, such oscillatory behavior does not take place. From Eq. (\ref{eq-run.3}) we can easily see that $|\ln F|$ does not take very large values unless $\tilde p$ is extremely small or large. For instance, for $\bar g=\bar g'=0.25$, $\ln F(10^{-50})\simeq 0.65$ and $\ln F(10^{50})\simeq -2.42$.
On the other hand, if $\bar g'=\beta \bar g$, $\frac{\bar g\bar g'}{15\bar g_+^2}$ is maximum at $\beta=1$, hence
$\frac{\bar g\bar g'}{15\bar g_+^2}\leq \frac{1}{30}$. Therefore, we can easily conclude that $\big\vert\frac{\bar g\bar g'}{15\bar g_+^2}\ln F\big\vert<\frac{\pi}{2}$ for typical values of $\bar g,\bar g'$, and $\tilde p$.
\begin{figure}
\centering
\begin{minipage}[t]{0.45\linewidth}
\centering
\includegraphics[scale=0.615]{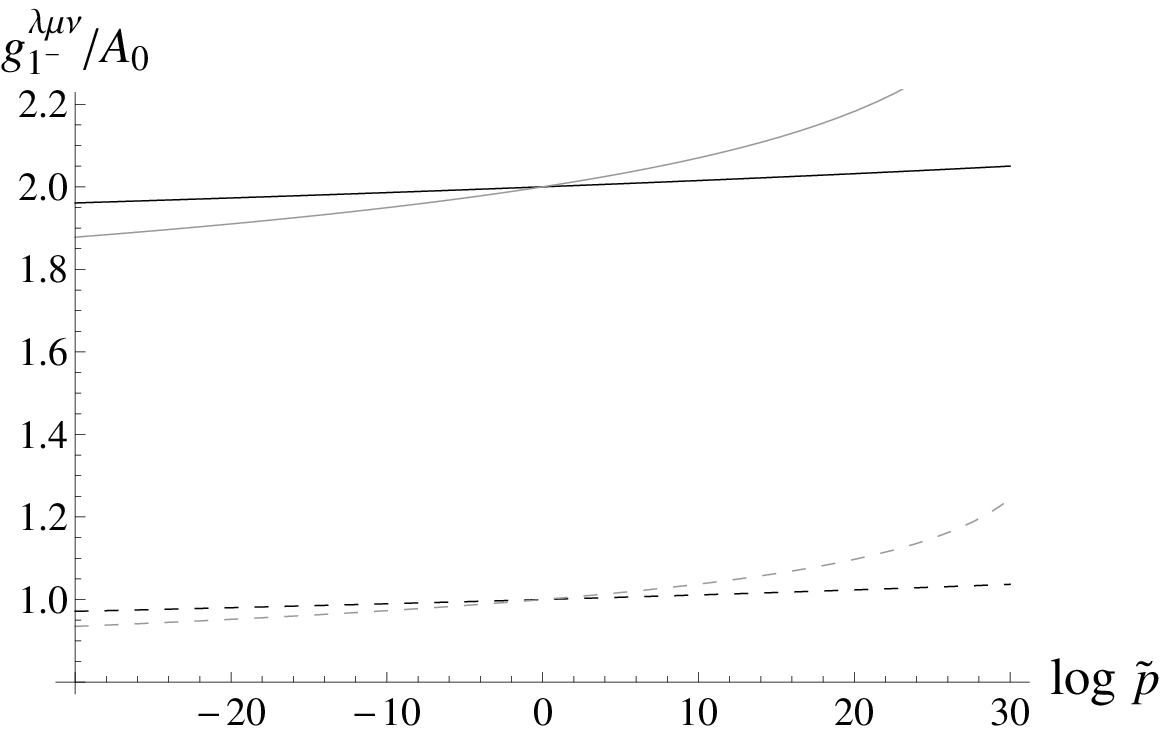}
\caption{$g^{\lambda\mu\nu}_{1^-}$ as a function of the scale. Here the black lines represent the conditions $2g=g'=0.2g_0$ and the gray lines the conditions $g=2g'=0.4g_0$. In the continuous lines $\bar g^{\lambda\mu\nu}_{1^-}=2\bar g^{\lambda\mu\nu}_{2^-}=2A_{0}$; in the dashed lines
 $\bar g^{\lambda\mu\nu}_{1^-}=\bar g^{\lambda\mu\nu}_{2^-}=A_{0}$.}.
\label{fig:4}
\end{minipage}
\hspace{0.5cm}
\begin{minipage}[t]{0.45\linewidth}
\centering
\includegraphics[scale=0.62]{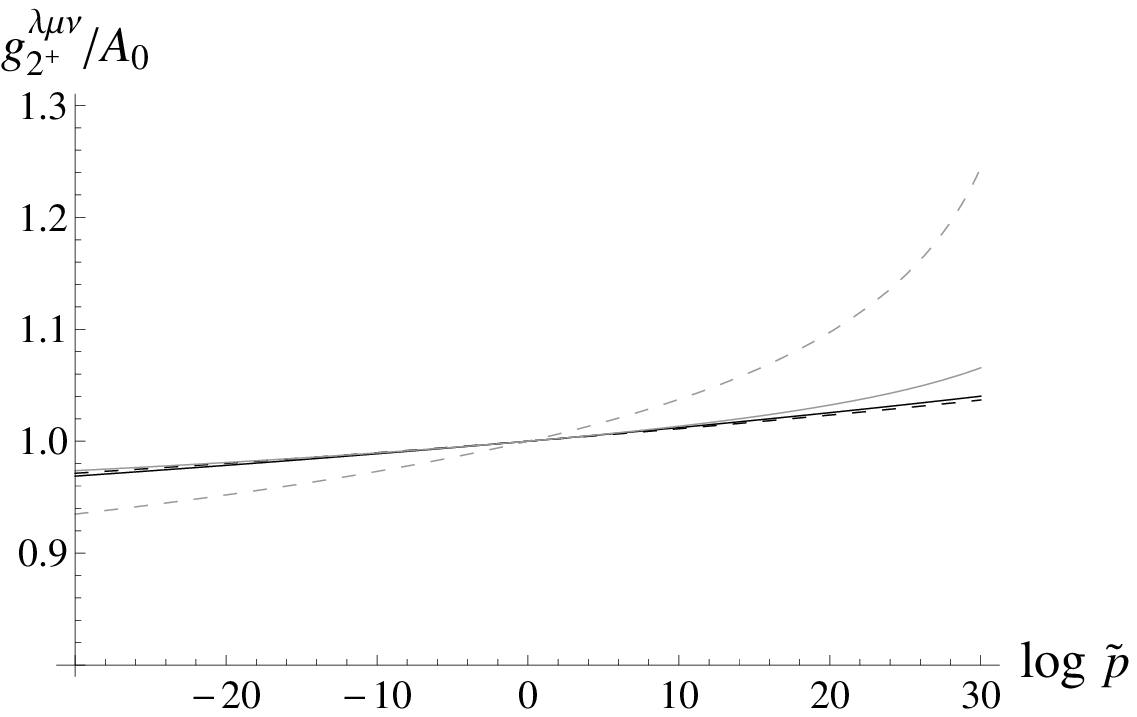}
\caption{$g_{2^+}^{\lambda\mu\nu}$ as a function of the scale. We used the same initial conditions and labeling as in figure \ref{fig:4}.}
\label{fig:5}
\end{minipage}
\end{figure}

\section{Application to Lorentz-violating QED}\label{sec:elec}

The method that we have used to decomposing into irreducible representations can also be applied to Lorentz-violating QED using the results from \cite{ref-kost3}.

In Lorentz invariant QED the evolution of the electron charge and its mass are given, respectively, by \cite{ref-kost3,ref-peskin}
\begin{align}
&e^{2}=\bar e^2Q^{-1}\,,\,\,\,\,m=\bar m Q^{9/4}\,,\,\,\,\,\,\textrm{with}\\
&Q(\tilde p)\equiv 1-(\bar e^2/6\pi^2)\ln\tilde p\,.
\end{align}
From \cite{ref-kost3} we can see that the parameters describing Lorentz-violating QED can also be decomposed into four groups, which are
\begin{enumerate}
\item[1.] Set $\{\eta^{\lambda\nu}c^{\,\mu\sigma},(k_F)^{\lambda\mu\nu\sigma}\}$.
\item[2.] Set $\{a^\mu,e^\mu,f^\mu\}$.
\item[3.] Set $\{d^{\,\mu\nu},H^{\mu\nu}\}$.
\item[4.] Set $\{ b^{\lambda}\eta^{\mu\nu},g^{\lambda\mu\nu},(k_{AF})^\lambda\eta^{\mu\nu}\}$.
\end{enumerate}

\subsection{Group 1}

In Lorentz-violating QED we also have the parameter $(k_F)_{\lambda\mu\nu\sigma}$ associated with the electromagnetic field. Since its contribution to the Lorentz-violating QED lagrange density is $\Delta\mathcal L=-\frac{1}{4}(k_F)_{\lambda\mu\nu\sigma}F^{\lambda\mu}F^{\nu\sigma}$, it has the symmetry properties \cite{ref-kost3}
\begin{subequations}
\begin{align}\label{sim1}
&(k_F)_{\lambda\mu\nu\sigma}=(k_F)_{\nu\sigma\lambda\mu}=-(k_F)_{\mu\lambda\nu\sigma}=-(k_F)_{\lambda\mu\sigma\nu}\,,\\\label{sim2}
&(k_F)_{\lambda\mu\nu\sigma}+(k_F)_{\lambda\nu\sigma\mu}+(k_F)_{\lambda\sigma\mu\nu}=0\,,\\\label{sim3}
&(k_F)_{\mu\nu}^{\phantom{\mu\nu}\mu\nu}=0\,.
\end{align}
\end{subequations}
From \cite{ref-kost3} we have the $\beta$-functions
\begin{subequations}
\begin{align}\label{betak-1}
(\beta_{k_F})^{\lambda\mu\nu\sigma}=&\,\frac{_{e^2}}{^{6\pi^2}}
\Big[(k_F)^{\lambda\mu\nu\sigma}\!-\eta^{[\lambda\nu}c_S^{\,\mu\sigma]}+\eta^{[\mu\nu}c_S^{\,\lambda\sigma]}
-\frac{_1}{^2}c^{\,\alpha}_{\,\,\,\alpha}\eta^{[\lambda\nu}\eta^{\mu\sigma]}\Big],\\\label{betak-1a}
%%%%%%%%%%%%%%%%%%%%%%%%%%%
(\beta_{c})^{\mu\sigma}=&\,\frac{_{e^2}}{^{6\pi^2}}
\Big[2c_S^{\,\mu\sigma}+\frac{_1}{^2}\eta^{\mu\sigma}c^{\,\alpha}_{\,\,\,\alpha}-(k_F)^{\alpha\mu\phantom{\nu}\sigma}_{\phantom{\mu\nu}\alpha}\Big].
\end{align}
\end{subequations}
where we used the definition $X^{[\lambda\nu}Y^{\mu\sigma]}\equiv X^{\lambda\nu}Y^{\mu\sigma}-X^{\mu\nu}Y^{\lambda\sigma}$.
Notice that previous relations satisfy the symmetry conditions stated in Eqs. (\ref{sim1})$-$(\ref{sim3}); these symmetry conditions also imply from Eq. (\ref{betak-1a}) that $c_A^{\,\mu\nu}=\bar c_A^{\,\mu\nu}$. From now on we will use the same definitions as in Eq. (\ref{def-traces}) and deal with $(k_F)^{\lambda\mu\nu\sigma}$ as explained in appendix \ref{appB}. Notice that when we decompose $(k_F)^{\lambda\mu\nu\sigma}$ according to Eq. (\ref{trace9}) we have permutations in the indices $\{\lambda,\mu,\nu,\sigma\}$ just like in Eq. (\ref{betak-1}). Hence, each permutation of indices satisfies the same equations and so a decomposition of the $\beta$-function into irreducible representations can be written as
\begin{subequations}
\begin{align}\label{betak-2}
(\beta_{k_F}^{\,t})^{\lambda\mu\nu\sigma}\!+\!\frac{_1}{^2}\eta^{\lambda\nu}(\beta_{k_F})_{2S}^{\mu\sigma}
&=\frac{_{e^2}}{^{6\pi^2}}\Big[(k_F)_t^{\lambda\mu\nu\sigma}+\frac{_1}{^2}\eta^{\lambda\nu}\bigl((k_F)_S^{\mu\sigma}-2c_S^{\,\mu\sigma}\bigr)
-\frac{_1}{^2}c^{\,\alpha}_{\,\,\,\alpha}\eta^{[\lambda\nu}\eta^{\mu\sigma]}\Big],\\
%%%%%%%%%%%%%%%%%%%%%%%%%%%
\frac{_1}{^4}\eta^{\mu\nu}(\beta_c)^{\alpha}_{\,\,\,\alpha}\!+\!(\beta_{c})^{\mu\sigma}_S&=\frac{_{e^2}}{^{6\pi^2}}
\Big[2c_S^{\,\mu\sigma}\!+\!\frac{_1}{^2}\eta^{\mu\sigma}\!c^{\,\alpha}_{\,\,\,\alpha}\!-(k_F)_{2S}^{\mu\sigma}\Big].
\end{align}
\end{subequations}
Taking the double trace of Eq. (\ref{betak-2}) we see that $c^{\,\alpha}_{\,\,\,\alpha}=0$. This is not inconsistent, however, because this parameter has no physical relevance because it can be absorbed by a redefinition in the fermion field \cite{ref-collad-4}. Multiplying Eq. (\ref{betak-2}) by $\eta_{\lambda\nu}$ we have the linear system
\begin{eqnarray}\label{betak-3}
\left(\!\!\!\begin{array}{c}(\beta_{k_F})_{2S}^{\mu\sigma}\\(\beta_{c})^{\mu\sigma}_S\end{array}\!\!\!\right)
=\frac{e^2}{6\pi^2}\left(\!\!\begin{array}{rr} 1 & -2 \\ -1 & 2\end{array}\!\!\right)\!\!
\left(\!\!\!\begin{array}{c}({k_F})_{2S}^{\mu\sigma}\\{c}^{\mu\sigma}_S\end{array}\!\!\right)\!.
\end{eqnarray}
The solutions are thus
\begin{subequations}
\begin{align}\label{betak-4}
(k_F)_{2S}^{\mu\sigma}&=(\bar k_F)_{2S}^{\mu\sigma}
-\frac{_1}{^3}\bigl((\bar k_F)_{2S}^{\mu\sigma}-2c_S^{\mu\sigma}\bigr)\bigl(1-Q^{-3}\bigr)\,,\\
c_{S}^{\mu\sigma}&=\bar c_{S}^{\,\mu\sigma}
+\frac{_1}{^3}\bigl((\bar k_F)_{2S}^{\mu\sigma}-2c_S^{\mu\sigma}\bigr)\bigl(1-Q^{-3}\bigr)\,.
\end{align}
\end{subequations}
Previous results imply that
\begin{subequations}
\begin{align}\label{betak-5}
%(\beta_{k_F}^{\,t})^{\lambda\mu\nu\sigma}=\frac{e^2}{6\pi^2}(k_F)_t^{\lambda\mu\nu\sigma},\,\,\,\,\textrm{so}\,\,\,\,
(k_F)_{t}^{\lambda\mu\nu\sigma}=(\bar k_F)_{t}^{\lambda\mu\nu\sigma}Q^{-1}\,.
\end{align}
\end{subequations}
Notice that the solutions found in Eqs. (\ref{betak-4}$-$\ref{betak-5}) are the same than those found in \cite{ref-kost3} once we write back each tensor field in terms of their irreducible representations.
%Nonetheless, the presented method is more convenient because the decoupling through irreducible representations reduces the complexity of the problem.

\subsection{Group 2}

As stated, vector parameters do not need to be decomposed into irreducible representations, so the decoupling is direct. From \cite{ref-kost3} we can easily see that $e^{\mu}(\tilde p)=\bar e^{\mu}$ and $f^{\mu}(\tilde p)=\bar f^{\mu}$. We use such results to integrate $a^\mu$; notice that this system cannot be solved through a matrix diagonalization because $(\beta_e)^\mu=(\beta_f)^\mu=0$ (the matrix would have two rows of zeroes). We confirmed the result \cite{ref-kost3}
\begin{eqnarray}
a^\mu\!\!\!&=&\!\!\!\bar a^\mu-\bar m\bar e^\mu\bigl(1-Q^{9/4}\bigr)\,.
\end{eqnarray}

\subsection{Group 3}

In order to find the solutions for this group we first decouple $d^{\,\mu\nu}$, whose $\beta$-function does not depend on $H^{\mu\nu}$ \cite{ref-kost3}. The solutions are
\begin{align}
d^{\,\alpha}_{\,\,\,\alpha}&=\bar d^{\,\alpha}_{\,\,\,\alpha}Q^{-2},\,\,\,\,
d^{\,\mu\nu}_S=\bar d^{\,\mu\nu}_SQ^{-2},\,\,\,\,
d_A^{\,\mu\nu}=\bar d_A^{\,\mu\nu}.
\end{align}
For $H^{\mu\nu}=H_A^{\mu\nu}$ we use the solution $d_A^{*\mu\nu}=\bar d_A^{*\mu\nu}$ to integrate $H_A^{\mu\nu}$. Like in the previous group, we have to follow this method instead of the matrix-diagonalization-procedure because $(\beta_d)_A^{\mu\nu}=0$; the integration gives
\begin{eqnarray}
H_A^{\mu\nu}=\bar H^{\mu\nu}_AQ^{-3/4}
+\bar m\bar d^{\,*\mu\nu}_AQ^{9/4}\bigl(1-Q^{-3}\bigr).
\end{eqnarray}
While the solution for $d^{\,\mu\nu}$ is the same, the solution for $H_A^{\mu\nu}$ differs from the one found in \cite{ref-kost3}; the reason is probably a typographical or algebraic mistake.

\subsection{Group 4}

We will first focus on the solutions for the coefficients $g^{\lambda\mu\nu}$ \cite{ref-kost3}. By permuting indices and decomposing into irreducible representations we find the $\beta$-functions for $g^{\lambda\mu\nu}$ and $g^{\lambda\nu\mu}$, which are, respectively
\begin{align}
(\beta_g)_A^{\lambda\mu\nu}+\frac{_4}{^3}(\beta_g)_{1^-}^{\lambda\mu\nu}
+\frac{_1}{^3}\eta^{\lambda[\nu]}(\beta_g)_2^{\mu}&=
\frac{_{e^2}}{^{8\pi^2}}\Big[\frac{_{16}}{^3}g_{1^-}^{\lambda\mu\nu}-\frac{_8}{^3}g_{2^+}^{\lambda\mu\nu}
+\frac{_4}{^3}\eta^{\lambda[\nu]}g_2^{\mu}+\frac{_5}{^3}\eta^{\lambda[\nu]}g_1^{\mu}\Big],\\
%%%%%%%%%%%%%%%%%%%%%%%%%%%%%%%%%%%%%%5
-(\beta_g)_A^{\lambda\mu\nu}+\frac{_4}{^3}(\beta_g)_{2^+}^{\lambda\mu\nu}
-\frac{_1}{^3}\eta^{[\lambda\nu}(\beta_g)_1^{\mu]}&=
\frac{_{e^2}}{^{8\pi^2}}\Big[\frac{_{4}}{^3}g_{2^+}^{\lambda\mu\nu}+\frac{_4}{^3}g_{1^-}^{\lambda\mu\nu}
+\frac{_1}{^3}\eta^{\lambda[\nu]}g_2^{\mu}+\frac{_1}{^6}\eta^{\lambda[\nu]}g_1^{\mu}\Big],
\end{align}
where we used again the condition $g_1^\mu+g_2^\mu+g_3^\mu=0$ (remember $X^{\lambda[\nu]\mu}\equiv X^{\lambda\nu\mu}-X^{\mu\nu\lambda}$).
If we multiply by $\varepsilon_{\lambda\mu\nu\sigma}$ in either equation we see that $(\beta_g^A)^{*\mu}=0$. Thus
$g_A^{\lambda\mu\nu}=\bar g_A^{\lambda\mu\nu}$ or $g_A^{*\mu}=\bar g_A^{*\mu}$.
Multiplying by $\eta_{\lambda\nu}$ in both equations we find the linear system
\begin{eqnarray}
\left(\!\!\begin{array}{c}(\beta_g)_1^\mu\\(\beta_g)_2^\mu\end{array}\!\!\right)=\frac{e^2}{8\pi^2}
\left(\!\!\begin{array}{rr}-\frac{1}{2} & -1\\ 5 & 4\end{array}\!\!\right)
\left(\!\!\begin{array}{c}g_1^\mu\\g_2^\mu\end{array}\!\!\right),
\end{eqnarray}
leading to the solutions
\begin{align}
g_1^{\mu}&=\bar g_1^{\mu}\bigl(5Q^{-9/8}\!-4Q^{-3/2}\bigr)
\!+2\bar g_2^\mu\bigl(Q^{-9/8}\!-Q^{-3/2}\bigr),\nonumber\\
g_2^{\mu}&=\bar g_2^{\mu}\bigl(5Q^{-3/2}\!-4Q^{-9/8}\bigr)
\!+10\bar g_1^\mu\bigl(Q^{-3/2}\!-Q^{-9/8}\bigr).
\end{align}
The scale dependence for the other rank-one coefficients is given by
\begin{subequations}
\begin{align}
(k_{AF})^{\mu}&=(\bar k_{AF})^{\mu}Q^{-1},
\\
%%%%%%%%%%%%%%%%%%%%%%
b^{\,\mu}&=\bar b^{\,\mu}
\!-\bar m\bar g_A^{*\mu}\bigl(1-Q^{9/4}\bigr)\!-\frac{_9}{^4}(\bar k_{AF})^{\mu}\!\bigl(1-Q^{-1}\bigr).
\end{align}
\end{subequations}
Additionally, using previous results we obtain the linear system
\begin{eqnarray}
\left(\!\!\begin{array}{c}(\beta_g)_{1^-}^{\lambda\mu\nu}\\(\beta_g)_{2^+}^{\lambda\mu\nu}\end{array}\!\!\right)
=\frac{e^2}{8\pi^2}\left(\!\!\begin{array}{rr}4 & -2\\ 1 & 1\end{array}\!\!\right)
\left(\!\!\begin{array}{c}g_{1^-}^{\lambda\mu\nu}\\g_{2^+}^{\lambda\mu\nu}\end{array}\!\!\right)\!,
\end{eqnarray}
whose solutions are
\begin{subequations}
\begin{align}
g_{1^-}^{\lambda\mu\nu}&=\bar g_{1^-}^{\lambda\mu\nu}\!\bigl(2Q^{-9/4}\!-Q^{-3/2}\bigr)
\!-2\bar g_{2^+}^{\lambda\mu\nu}\!\bigl(Q^{-9/4}\!-Q^{-3/2}\bigr),\\
%%%%%%%%%%%%%%%%%%%%%
g_{2^+}^{\lambda\mu\nu}&=\bar g_{2^+}^{\lambda\mu\nu}\!\bigl(2Q^{-3/2}\!-Q^{-9/4}\bigr)
\!-\bar g_{1^-}^{\lambda\mu\nu}\!\bigl(Q^{-3/2}\!-Q^{-9/4}\bigr).
\end{align}
\end{subequations}

\subsubsection{Inclusion of Yukawa terms in $g^{\lambda\mu\nu}$}
\begin{figure}
\centering
\includegraphics[scale=0.6]{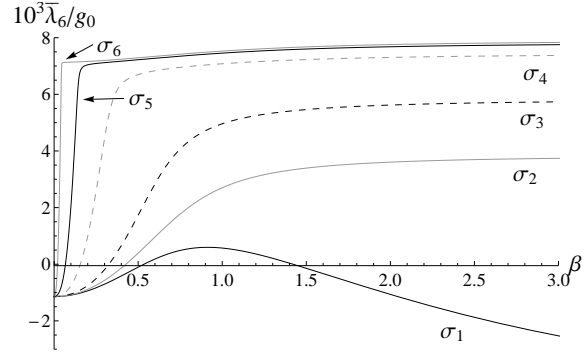}
\caption{$\bar\lambda_6$ as a function of $\bar g$, $\bar g'$, and $\bar e$. Here we have $\bar g=g_0$, $\bar g'=\beta g_0$, and $\bar e=\sigma g_0$. For each one of the six plots we have $\sigma_1=0.25$, $\sigma_2=0.6$, $\sigma_3=1$, $\sigma_4=2.5$, $\sigma_5=6$, and $\sigma_6=20$. It can easily be seen that when electromagnetic effects are included in Eq. (\ref{eq-mat0}), the negative eigenvalue tends to become more positive. This effect is larger for larger $\sigma$.}
\label{fig:6}
\end{figure}
Let us remember that Eq. (\ref{eq-mat0}) described the set of differential equations for the parameters $b^\mu,J^\mu,g_1^\mu,g_2^\mu$, and $g_A^{*\mu}$ in the pure Yukawa regime. When we also include electromagnetic contributions we will find a linear system of the form $(\beta_x)_m^\mu=\sum_{m=1}^6M_{mn}x_{n}^\mu$, with
$(\vec {x}^{\,T})^\mu=(k_{AF},b,J,g_1,g_2,g_A^*)^\mu$ and
\begin{align}\label{eq-mat01}
M=\eta\left(\!\!\begin{array}{cccccc}
\frac{8}{3}e^2 \!\!&\!\! 0 \!\!&\!\! 0 \!\!&\!\! 0 \!\!&\!\! 0 \!\!&\!\! 0 \\
6e^2 \!\!&\!\! g_+^2 \!\!&\!\! 2(mg\!+\!m'g') \!\!&\!\! 0 \!\!&\!\! \frac{1}{2}m'g_+^2 \!\!&\!\! \frac{3}{2}mg_+^2\!-6me^2\\
0 \!\!&\!\! 0 \!\!&\!\! 8g_+^{2} \!\!&\!\! 0 \!\!&\!\! g'g_+^2 \!\!&\!\! 3gg_+^2\\
0 \!\!&\!\! 0 \!\!&\!\! -3g' \!\!&\!\! \frac{2}{3}(2g^2\!+\!g'^{\,2})\!-e^2 \!\!&\!\! \frac{7}{6}g_-^2-2e^2 \!\!&\!\! -3gg'\\
0 \!\!&\!\! 0 \!\!&\!\! 6g' \!\!&\!\! \frac{2}{3}g_-^2+10e^2 \!\!&\!\! \frac{1}{3}(g^2\!+\!5g'^{\,2})\!+\!8e^2 \!\!&\!\! 6gg'\\
0 \!\!&\!\! 0 \!\!&\!\! 2g \!\!&\!\! \frac{4}{9}gg' \!\!&\!\! \frac{8}{9}gg' \!\!&\!\! 2g^2
\end{array}\!\!\!\right).
\end{align}
Last matrix has 6 different eigenvalues. While two or them can be read automatically from last matrix and are $\lambda_1=\frac{8\eta}{3}e^2$ and $\lambda_2=\eta g_+^2$, the remaining four take very complicated forms. Nonetheless, three of them, name them $\lambda_3,\lambda_4$ and $\lambda_5$ are always positive for any combination of $\bar g$, $\bar g'$, and $\bar e$. $\lambda_6$ is not always positive for any combination of such parameters, which can be seen in figure \ref{fig:6}.

\section{Summary and Conclusions}\label{sec:conc}

We have used a method based on decomposing a tensor into its irreducible representations to solve the $\beta$-functions of two theories with Lorentz and CPT violations. Although we mainly focused on a Yukawa theory because the scale dependence of its coefficients had not been solved, we also applied this method to an electromagnetic theory for comparison reasons. Indeed, up to some possible algebraic or typographical mistakes we confirmed the results found in \cite{ref-kost3}. Nonetheless, our method is more convenient because it reduces the complexity of the problem.

As stated, those operators whose scale dependence is described by negative $\beta$-functions could lead to an inconsistent theory if their scale dependence is strong enough, thus predicting a large imprint of Lorentz violation at low energies, which, at best of our knowledge, has not been observed. Fortunately, in their respective diagonal bases, most coefficients are described by positive $\beta$-functions. In this way we could envision, within the scope of these models, Lorentz symmetry as an emergent low-energy property.

Some mass operators have negative $\beta$-functions; however, their scale dependence is weak enough and so their imprint of LV at low energies is negligible. This fact also took place in some of the irreducible representations of $g^{\lambda\mu\nu}$. Since the solutions obtained for the analyzed Yukawa theory do not predict large imprints of Lorentz violation at low enough energies, this model is consistent with observations, at least, at one-loop order. Actually, the values of the couplings and masses of the Lorentz invariant theory are not constrained by this condition.

This method could also be applied to renormalize other Lorentz-violating theories with similar characteristics or any other theory that has similar properties. Studying the scale behavior of a Lorentz-violating QCD theory could be of particular interest because, in the Lorentz invariant case, it is asymptotically free.

\appendix

\vspace{1cm}

\hspace{-0.6cm}{\LARGE{\bf Appendices}}

\section{Irreducible decomposition of three-rank tensors}\label{appAa}

The permutations of three indices form the group $S_3$. The irreducible representations of three-rank tensors can be found using the following set of Young diagrams
\begin{center}
\includegraphics[scale=0.9]{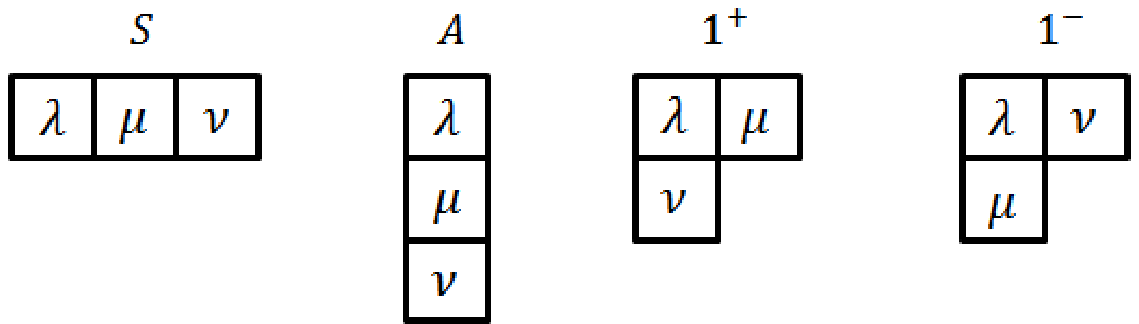}
\end{center}
Last set of diagrams generates the four irreducible representations
\begin{subequations}
\begin{align}
X_S^{\lambda\mu\nu}&=
\frac{_1}{^6}\bigl(X^{(\lambda\mu)\nu}+X^{\lambda(\nu)\mu}+X^{\nu(\lambda\mu)}\bigr),\\
X_A^{\lambda\mu\nu}&=
\frac{_1}{^6}\bigl(X^{[\lambda\mu]\nu}+X^{\mu[\nu]\lambda}+X^{[\nu\lambda]\mu}\bigr),\\
X_{1^+}^{\lambda\mu\nu}&=\frac{_1}{^4}\bigl(X^{(\lambda\mu)\nu}-X^{\nu(\mu\lambda)}\bigr),\\
X_{1^-}^{\lambda\mu\nu}&=\frac{_1}{^4}\bigl(X^{[\lambda\mu]\nu}+X^{[\nu\mu]\lambda}\bigr),
\end{align}
\end{subequations}
with $X^{(\lambda\mu)\nu}\equiv X^{\lambda\mu\nu}+X^{\mu\lambda\nu}$,
$X^{[\lambda\mu]\nu}\equiv X^{\lambda\mu\nu}-X^{\mu\lambda\nu}$, $X^{\lambda(\nu)\mu}\equiv X^{\lambda\nu\mu}+X^{\mu\nu\lambda}$, 
and $X^{\lambda[\nu]\mu}\equiv X^{\lambda\nu\mu}-X^{\mu\nu\lambda}$.
However, last set only generates $4$ linearly-independent representations. In order to complete six independent combinations, we will exchange the indices $\mu\rightarrow\nu$ in the tensors generated by last two diagrams. So we define
\begin{equation}
X_{2^+}^{\lambda\mu\nu}=X_{1^+}^{\lambda\nu\mu}\,,\,\,\,\,\,X_{2^-}^{\lambda\mu\nu}=X_{1^-}^{\lambda\nu\mu}\,.
\end{equation}
This set of six tensors can be inverted to find the relations
\begin{equation}\label{eq:transf}
\left(\!\!\!\begin{array}{c}X^{\lambda\mu\nu}\\X^{\lambda\nu\mu}\\X^{\mu\nu\lambda}\\
X^{\mu\lambda\nu}\\X^{\nu\lambda\mu}\\X^{\nu\mu\lambda}\end{array}\!\!\!\right)=\frac{1}{3}
\left(\!\!\begin{array}{rrrrrr}
3 \!\!&\!\! 3  \!\!&\!\! 4  \!\!&\!\! 0  \!\!&\!\! 4  \!\!&\!\! 0 \\ 
3 \!\!&\!\! -3 \!\!&\!\! 0  \!\!&\!\! 4  \!\!&\!\! 0  \!\!&\!\! 4 \\ 
3 \!\!&\!\! 3  \!\!&\!\! -4 \!\!&\!\! -4 \!\!&\!\! 0  \!\!&\!\! 4 \\
3 \!\!&\!\! -3 \!\!&\!\! 4  \!\!&\!\! 0  \!\!&\!\! -4 \!\!&\!\! -4 \\ 
3 \!\!&\!\! 3  \!\!&\!\! 0  \!\!&\!\! 4  \!\!&\!\! -4 \!\!&\!\! -4 \\ 
3 \!\!&\!\! -3 \!\!&\!\! -4 \!\!&\!\! -4 \!\!&\!\! 4  \!\!&\!\! 0 \\
\end{array}\!\!\right)\!\!
\left(\!\!\begin{array}{c}X_S^{\lambda\mu\nu}\\X_A^{\lambda\mu\nu}\\X_{1^+}^{\lambda\mu\nu}\\
X_{2^+}^{\lambda\mu\nu}\\X_{1^-}^{\lambda\mu\nu}\\X_{2^-}^{\lambda\mu\nu}\end{array}\!\!\right)\!.
\end{equation}
We can verify that $X^{\!*\lambda\mu\nu}=X^{*\lambda\mu\nu}_A$ as well as the following relations
\begin{subequations}
\begin{align}
X_{1^+}^{\mu\lambda\nu}&=X_{1^+}^{\lambda\mu\nu},\,\,\,\,\,
X_{1^+}^{\lambda\nu\mu}=X_{1^+}^{\nu\lambda\mu}=X_{2^+}^{\lambda\mu\nu},
\nonumber\\
X_{1^+}^{\nu\mu\lambda}&=X_{1^+}^{\mu\nu\lambda}=-\,(X_{1^+}^{\lambda\mu\nu}+X_{2^+}^{\lambda\mu\nu}),
\\
X_{1^-}^{\nu\mu\lambda}&=X_{1^-}^{\lambda\mu\nu},\,\,\,\,\,
X_{1^-}^{\lambda\nu\mu}=X_{1^-}^{\mu\nu\lambda}=X_{2^-}^{\lambda\mu\nu},
\nonumber\\
X_{1^-}^{\mu\lambda\nu}&=X_{1^-}^{\nu\lambda\mu}=-\,(X_{1^-}^{\lambda\mu\nu}+X_{2^-}^{\lambda\mu\nu}).
\end{align}
\end{subequations}
Now we want to isolate the ``trace-like" components, which have different transformation rules. Let us define
\begin{eqnarray}\label{traces3}
X_1^\sigma\equiv X^{\sigma\alpha}_{\phantom{\nu\alpha}\alpha}\,,\,\,\,
X_2^\sigma\equiv X^{\alpha\sigma}_{\phantom{\alpha\nu}\alpha}\,,\,\,\,
X_3^\sigma\equiv X^{\alpha\phantom{\alpha}\sigma}_{\phantom{\alpha}\alpha}\,.
\end{eqnarray}
In order to isolate these contributions we must make the six irreducible tensors previously defined traceless in all of their components. So we make
\begin{subequations}
\begin{align}\label{eq:X1}
X_{St}^{\lambda\mu\nu}&=X_S^{\lambda\mu\nu}
-\frac{_1}{^{18}}\eta^{\lambda\mu}\bigl(X_{1}+X_{2}\!+\!X_{3}\bigr)^{\nu}
-\frac{_1}{^{18}}\eta^{\lambda(\nu)}\bigl(X_{1}+X_{2}+X_{3}\bigr)^{\mu},\\
X_{At}^{\lambda\mu\nu}&=X_A^{\lambda\mu\nu},\\
X_{1t^{+}}^{\lambda\mu\nu}&=X_{1^+}^{\lambda\mu\nu}
\!+\!\frac{_1}{^{6}}\eta^{\lambda\mu}\!\bigl(X_{1}-X_{3}\bigr)^{\nu}
\!-\!\frac{_1}{^{12}}\eta^{\lambda(\nu)}\!\bigl(X_{1}-X_{3}\bigr)^{\mu},\\ \label{eq:X4}
X_{1t^{-}}^{\lambda\mu\nu}&=X_{1^-}^{\lambda\mu\nu}
\!+\!\frac{_1}{^{6}}\eta^{\lambda\nu}\!\bigl(X_{1}-X_{2}\bigr)^{\mu}
\!-\!\frac{_1}{^{12}}\eta^{\lambda(\mu)}\!\bigl(X_{1}-X_{2}\bigr)^{\nu},
\end{align}
\end{subequations}
where the subindex ``$t$" means that the tensors are traceless and, again, $X_{2t^{+}}^{\lambda\mu\nu}=X_{1t^{+}}^{\lambda\nu\mu}$ and
$X_{2t^{-}}^{\lambda\mu\nu}=X_{1t^{-}}^{\lambda\nu\mu}$. The traceless representations of $X^{\lambda\mu\nu}_t$ (and its permutations) can be found using Eq. (\ref{eq:transf}).

From now on we will drop the traceless notation ``$t$'' and assume that the irreducible tensors in Eqs. (\ref{eq:X1})$-$ (\ref{eq:X4}) are traceless.

For our particular situation of interest, we will focus on the properties of $X^{\lambda\mu\nu}$ when it is antisymmetric under the exchange $\lambda\leftrightarrow\mu$ (this is the symmetry condition obeyed by $g^{\lambda\mu\nu}$ as introduced in Eq. (\ref{yuk.2})). This means that we can write $X^{\lambda\mu\nu}$ as $X^{\lambda\mu\nu}=\frac{1}{2}X^{[\lambda\mu]\nu}$. This condition reduces the number of linearly-independent irreducible representations (neglecting the ``traces") from six to three, which now are
\begin{subequations}
\begin{align}
X_{A}^{[\lambda\mu]\nu}&=\frac{_1}{^6}\bigl(X^{[\lambda\mu]\nu}+X^{\mu[\nu]\lambda}+X^{[\nu\lambda]\mu}\bigr),\label{eq-ga}\\
X_{2^+}^{[\lambda\mu]\nu}&=\frac{_1}{^4}\bigl(X^{\lambda[\nu]\mu}-\frac{_1}{^2}X^{[\mu\lambda]\nu}
+\frac{_1}{^2}X^{\nu[\lambda\mu]}+\frac{_1}{^2}\eta^{\lambda[\nu]}\!\left(X_{1}\!-\!X_{3}\right)^{\mu}\bigr),\\
X_{1^-}^{[\lambda\mu]\nu}&=\frac{_1}{^4}\big(X^{[\lambda\mu]\nu}
-\frac{_1}{^2}X^{\mu[\nu]\lambda}+\frac{_1}{^2}X^{\nu[\mu\lambda]}+\frac{_1}{^2}\eta^{\lambda[\nu]}(X_{1}-X_{2})^{\mu}\big).
\end{align}
\end{subequations}
Notice that $\eta_{\lambda\mu}X^{[\lambda\mu]\nu}=0$, $\eta_{\lambda\nu}X^{[\lambda\mu]\nu}=\frac{1}{2}(X_2-X_1)^\mu$, and
$\eta_{\mu\nu}X^{[\lambda\mu]\nu}=\frac{1}{2}(X_1-X_2)^\lambda$. We will now use the definitions
$X_{1}^{[\alpha]}\equiv\frac{1}{2}(X_1^\alpha-X_3^\alpha)$, $X_{2}^{[\alpha]}\equiv\frac{1}{2}(X_2^\alpha-X_1^\alpha)$,
and $X_{3}^{[\alpha]}\equiv\frac{1}{2}(X_3^\alpha-X_2^\alpha)$. By construction we have the property
\begin{eqnarray}\label{equices}
X_{1}^{[\alpha]}+X_{2}^{[\alpha]}+X_{3}^{[\alpha]}=0\,.
\end{eqnarray}
In order to simplify the notation we will assume from now on that $X^{\lambda\mu\nu}$ and its permutations are antisymmetric under the exchange $\lambda\leftrightarrow\mu$ and
so $X^{[\lambda\mu]\nu}\rightarrow X^{\lambda\mu\nu}$. Relabeling $X_{i}^{[\alpha]}\rightarrow X_i^\alpha$ we find
\begin{subequations}
\begin{align}\label{gestil1}
X^{\lambda\mu\nu}&=X_A^{\lambda\mu\nu}
+\frac{_4}{^3}X_{1^-}^{\lambda\mu\nu}+\frac{_1}{^{3}}\eta^{\lambda[\nu]}X_2^{\mu}
%\,,\,\,\,\,\textrm{with}\,\,\,\eta_{\mu\nu}X^{\lambda\mu\nu}=-X_2^\lambda\,
,
\\
%%%%%%%%%%%%%%%%%%%%%%%%
X^{\lambda\nu\mu}&=-X_A^{\lambda\mu\nu}+\frac{_4}{^3}X_{2^+}^{\lambda\mu\nu}
-\frac{_1}{^{3}}\eta^{\lambda[\nu]}X_1^{\mu}
%\,,\,\,\,\,\textrm{with}\,\,\,\eta_{\lambda\nu}X^{\lambda\nu\mu}=-X_1^\mu\,
,
\\
%%%%%%%%%%%%%%%%%%%%%%%%
X^{\nu\mu\lambda}&=-X_A^{\lambda\mu\nu}-\frac{_4}{^3}X_{2^+}^{\lambda\mu\nu}+\frac{_4}{^3}X_{1^-}^{\lambda\mu\nu}
-\frac{_1}{^{3}}\eta^{\lambda[\nu]} X_3^{\mu}
%\,,\,\,\,\,\textrm{with}\,\,\,\eta_{\lambda\nu}X^{\nu\mu\lambda}=-X_3^\mu\,
,\\\label{gestil6}
%%%%%%%%%%%%%%%%%%%%%%%%%%%%%%%%%
X^{\mu\nu\lambda}&=-X^{\lambda\nu\mu},\,\,
X^{\mu\lambda\nu}=-X^{\lambda\mu\nu},\,\,
X^{\nu\lambda\mu}=-X^{\nu\mu\lambda},
%%%%%%%%%%%%%%%%%%%%%%%%%%%%%%%%
\end{align}
\end{subequations}
where $X_i^\alpha$ satisfy Eq. (\ref{equices}).

\section{Irreducible decomposition of four-rank tensors}\label{appB}

A four-rank tensor $X^{\lambda\mu\nu\sigma}$ must also be decomposed into irreducible representations under the group $S_4$; its associated Young diagrams are
\begin{center}
\includegraphics[scale=1]{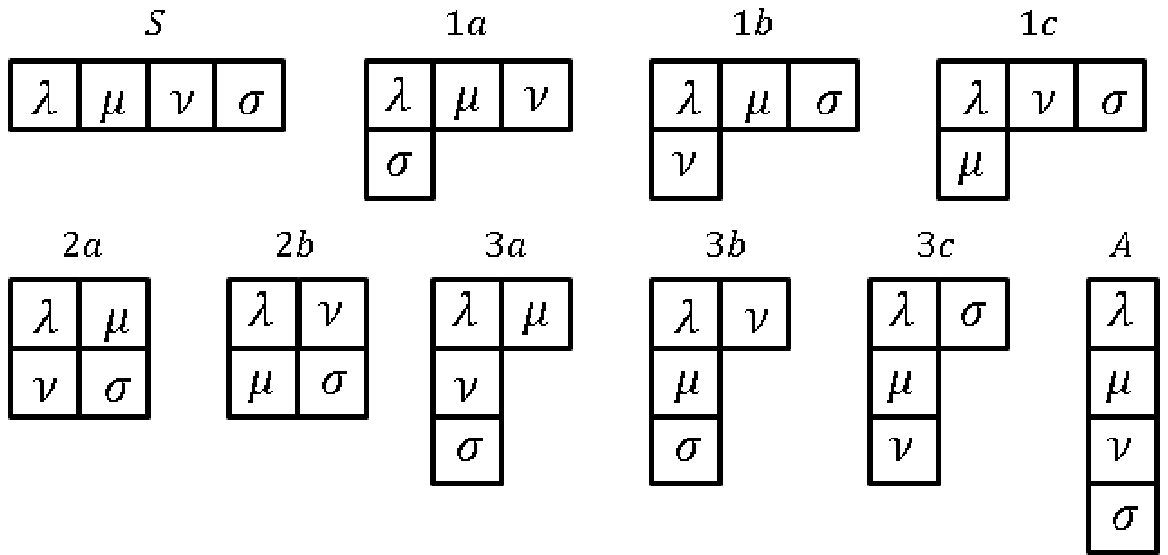}
\end{center}
and permutations of the indices $\{\lambda,\mu,\nu,\sigma\}$ to complete 24 different representations. In terms of the set of Young diagrams we can write
\begin{align}
X^{\lambda\mu\nu\sigma}=&\,
x_1X_S^{\lambda\mu\nu\sigma}\!+x_2X_{1a}^{\lambda\mu\nu\sigma}\!+x_3X_{1b}^{\lambda\mu\nu\sigma}
\!+x_4X_{1c}^{\lambda\mu\nu\sigma}
+x_5X_{2a}^{\lambda\mu\nu\sigma}+x_6X_{2b}^{\lambda\mu\nu\sigma}
\nonumber\\
&+x_7X_{3a}^{\lambda\mu\nu\sigma}+x_8X_{3b}^{\lambda\mu\nu\sigma}
+x_9X_{3c}^{\lambda\mu\nu\sigma}+x_{10}X_A^{\lambda\mu\nu\sigma},
\end{align}
where $x_i$ are rational numbers.
Fortunately, the $\beta$-functions of Lorentz violating QED that deal with four-rank tensors only include the combinations $(k_F)_{\lambda\mu\nu\sigma}$ and $(k_F)_{\lambda\alpha\nu}^{\phantom{1111}\alpha}$. Hence, we do not need to find all irreducible representations, we just need to focus on the trace and double trace representations of $(k_F)_{\lambda\mu\nu\sigma}$. Let us define
\begin{align}\label{def-traces}
X_1^{\alpha\beta}=&\, X^{\rho\phantom{\rho}\alpha\beta}_{\phantom{\rho}\rho}\,,\,\,\,
X_2^{\alpha\beta}\equiv X^{\rho\alpha\phantom{\rho}\beta}_{\phantom{\rho\alpha}\rho}\,,\,\,\,
X_3^{\alpha\beta}\equiv X^{\rho\alpha\beta}_{\phantom{\rho\alpha\beta}\rho},\nonumber\\
%%%%%%%%%%%%%%%%%%%%%
X_4^{\alpha\beta}=&\, X^{\alpha\rho\beta}_{\phantom{\rho\alpha\beta}\rho}\,,\,\,\,
X_5^{\alpha\beta}\equiv X^{\alpha\rho\phantom{\rho}\beta}_{\phantom{\rho\alpha}\rho}\,,\,\,\,
X_6^{\alpha\beta}\equiv X^{\alpha\beta\rho}_{\phantom{\alpha\beta\rho}\rho},\nonumber\\
%%%%%%%%%%%%%%%
X_7=&\, X^{\rho\phantom{\rho}\rho'}_{\phantom{\rho}\rho\phantom{\rho}\rho'}\,,\,\,\,
X_8\equiv X^{\rho\rho'\phantom{\rho\rho'}}_{\phantom{\rho\rho'}\!\!\rho\rho'}\,,\,\,\,
X_9\equiv X^{\rho\rho'\phantom{\rho\rho'}}_{\phantom{\rho\rho'}\!\!\rho'\rho}\!.
\end{align}
In terms of these definitions we can write the traceless and double traceless tensor $X^{\lambda\mu\nu\sigma}_t$ as
\begin{align}
&X^{\lambda\mu\nu\sigma}_t=X^{\lambda\mu\nu\sigma}\nonumber\\
&+\frac{_1}{^{96}}\eta^{\lambda\mu}\Big[\!-31X_1\!+\!X_1^T
\!+\!7X_2\!-\!X_2^{T}\!-\!X_3\!+\!7X_3^{T}\!-\!X_4\!+\!7X_4^{T}\!+\!7X_5\!-\!X_5^{T}\!-\!3X_6\!-\!3X_6^{T}
\Big]^{\nu\sigma}
%%%%%%%%
\nonumber\\
&+\frac{_1}{^{96}}\eta^{\lambda\nu}\Big[7X_1\!-\!X_1^T\!-\!31X_2\!+\!X_2^{T}\!+\!7X_3\!-\!X_3^{T}\!
-\!3X_4\!-\!3X_4^{T}\!+\!7X_5\!-\!X_5^{T}\!-\!X_6\!+\!7X_6^{T}\Big]^{\mu\sigma}
%%%%%%%%%%
\nonumber\\
&+\frac{_1}{^{96}}\eta^{\lambda\sigma}\Big[\!-X_1\!+\!7X_1^T\!+\!7X_2\!-\!X_2^{T}\!-\!31X_3\!+\!X_3^{T}\!
+\!7X_4\!-\!X_4^{T}\!-\!3X_5\!-\!3X_5^{T}\!-\!X_6\!+\!7X_6^{T}\Big]^{\mu\nu}
%%%%%%%%%%
\nonumber\\
&+\frac{_1}{^{96}}\eta^{\mu\nu}\Big[7X_1\!-\!X_1^T
\!+\!7X_2\!-\!X_2^{T}\!-\!3X_3\!-\!3X_3^{T}\!+\!7X_4\!-\!X_4^{T}\!-\!31X_5\!+\!X_5^{T}\!+\!7X_6\!-\!X_6^{T}
\Big]^{\lambda\sigma}\,\,\,
%%%%%%%%%%
\nonumber\\
&+\frac{_1}{^{96}}\eta^{\mu\sigma}\Big[\!-X_1\!+\!7X_1^T
\!-\!3X_2\!-\!3X_2^{T}\!+\!7X_3\!-\!X_3^{T}\!-\!31X_4\!+\!X_4^{T}\!+\!7X_5\!-\!X_5^{T}\!+\!7X_6\!-\!X_6^{T}
\Big]^{\lambda\nu}\,\,\,
%%%%%%%%%%
\nonumber\\
&+\frac{_1}{^{96}}\eta^{\nu\sigma}\Big[\!-3X_1\!-\!3X_1^T
\!-\!X_2\!+\!7X_2^{T}\!-\!X_3\!+\!7X_3^{T}\!+\!7X_4\!-\!X_4^{T}\!+\!7X_5\!-\!X_5^{T}\!-\!31X_6\!+\!X_6^{T}
\Big]^{\lambda\mu}\,\,\,
%%%%%%%%%%
\nonumber\\
&+\frac{_1}{^{144}}\eta^{\lambda\mu}\eta^{\nu\sigma}\Big[17X_7-7X_8-7X_9\Big]
+\frac{_1}{^{144}}\eta^{\lambda\nu}\eta^{\mu\sigma}\Big[-7X_7+17X_8-7X_9\Big]
%%%%%%%%%%
\nonumber\\
&+\frac{_1}{^{144}}\eta^{\lambda\sigma}\eta^{\mu\nu}\Big[-7X_7-7X_8+17X_9\Big],
\end{align}
where $(X^{T})^{\alpha\beta}=X^{\beta\alpha}$.

We will now study the properties of four-rank tensors obeying Eqs. (\ref{sim1})$-$(\ref{sim3}).

When we multiply by $\eta_{\lambda\nu}$, $\eta_{\mu\sigma}$, and $\eta_{\lambda\nu}\eta_{\mu\sigma}$ respectively in Eq. (\ref{sim2}) we obtain the conditions
\begin{subequations}
\begin{align}
(X_1)_S^{\mu\sigma}+(X_2)_S^{\mu\sigma}+(X_3)_S^{\mu\sigma}&=0\,,\\
(X_4)_S^{\lambda\nu}+(X_5)_S^{\lambda\nu}+(X_6)_S^{\lambda\nu}&=0\,,\\
X_7+X_8+X_9&=0\,.%\textrm{so}\,\,\,\,
\end{align}
\end{subequations}
Eq. (\ref{sim3}) implies $X_8=0$, so $ X_7=-X_9$. However, $X_7=X^{\lambda\mu\nu\sigma}\eta_{\lambda\mu}\eta_{\nu\sigma}=0$ because of the antisymmetry properties in Eq. (\ref{sim1}), so $X_9$ also vanishes. This implies that $X^{\lambda\mu\nu\sigma}$ is double traceless in all components.

Using the symmetry properties stated in Eq. (\ref{sim1}) we find that $X^{\lambda\mu\nu\sigma}$ can be simplified to
\begin{align}\label{trac2}
X^{\lambda\mu\nu\sigma}=&\,X^{\lambda\mu\nu\sigma}_t
+\frac{_1}{^{8}}\eta^{\lambda\nu}(X_a)_S^{\mu\sigma}
-\frac{_1}{^{8}}\eta^{\mu\nu}(X_a)_S^{\lambda\sigma}
-\frac{_1}{^{8}}\eta^{\lambda\sigma}(X_a)_S^{\mu\nu}
+\frac{_1}{^{8}}\eta^{\mu\sigma}(X_a)_S^{\lambda\nu}\,,
\end{align}
where $X_a\equiv X_2-X_3+X_4-X_5$.

Notice that the symmetry conditions obeyed by $X^{\lambda\mu\nu\sigma}$ also imply that
$(X_2)_S^{\mu\sigma}=(X_4)_S^{\mu\sigma}=
-(X_3)_S^{\mu\sigma}=-(X_5)_S^{\mu\sigma}=\frac{1}{4}(X_a)_S^{\mu\sigma}$, so we finally have
\begin{align}\label{trace9}
X^{\lambda\mu\nu\sigma}=&\,X^{\lambda\mu\nu\sigma}_t
+\frac{_1}{^{2}}\eta^{\lambda\nu}(X_2)_S^{\mu\sigma}
-\frac{_1}{^{2}}\eta^{\mu\nu}(X_2)_S^{\lambda\sigma}
-\frac{_1}{^{2}}\eta^{\lambda\sigma}(X_2)_S^{\mu\nu}
+\frac{_1}{^{2}}\eta^{\mu\sigma}(X_2)_S^{\lambda\nu}\,,
\end{align}
which is enough to decompose $(k_F)_{\lambda\mu\nu\sigma}$.

\end{document}